\documentclass[12pt]{iopart}\usepackage{iopams}

\usepackage{graphicx}%nova org
\usepackage{epsfig}%nova org
\usepackage{eucal} %jxzj
\usepackage{amssymb}%nova org
\usepackage{amsfonts}%nova org
\usepackage{amsbsy}%jxzj uncommond
\usepackage{mathrsfs} % mathscr

\usepackage{mathrsfs}
\def\subempty{\empty}

\def\strvac{\supvac}
\def\supvac{0}
\def\strempt{{}}

\def\ArticleLabel-lp{16}

\def\vir{{\rm vir}}

\def\RefUnif{3}

\def\citeUnif{3}
\def\vel{\upsilon}
\def\Eb{{\bf{E}}}
\def\Bb{{\bf{B}}}
\def\obs{{\rm obs}}
\def\ev{\epsilon}

\def\ke{\kappa}

\def\Omegavel{\mathbin{{\mit\Omega}\mkern-13.mu^{_{\mbox{$-$}}}\hspace{-0.08cm}{}_d }}

\def\empty{{\mbox{\tiny${\emptyset}$}}}

\def\Kcal{{\math{K}}}

\def\imr{{\rm im}}

\def\Thm{\vartheta}

\def\Xima{\xi}
\def\Vel{W}

\def\Pw{{\mathcal{P}}}

\def\Mcal{{\mathfrak{M}}}
\def\nablab{{\pmb{\nabla}}}
\def\velb{{\bf{v}}}

\def\jb{{\bf{j}}}
\def\fb{{\bf{f}}}
\def\Fb{{\bf{F}}}

\def\med{{\med}}

\def\Jcal{j}

\def\Jobs{j_{\rm obs}}

\def\obs{{\rm{obs}} }
\def\qm{{\rm qm}}
\begin{document}
%\baselineskip 0.46cm

%\def\PaperNumber{14}
%\renewcommand{\PaperNumber}{001}

%    \FirstPageHeading
%    \ShortArticleName{Doebner-Goldin  Equation based on Solutions for an Electrodynamic Model Particle}
%    \ArticleName{Motivation for Doebner-Goldin Equation based on Solutions for an Electrodynamic Model Particle. The Implied Applications}
%     \Author{J.X. Zheng-Johansson$^{\dag, \ddag}$ }
%     \Address{$^\dag$. Institute of Fundamental Physics Research,  611 93 Nyk\"oping, Sweden. \\ $^\ddag$. In affiliation with the Swedish Institute of  Space Physics, Kiruna, Sweden }
%\emailD{jxzj@iofpr.org} % E-mail address of First Author
%\URLaddressD{http://iofpr.org/\~{}myHome/} %URL address of First Author

%\title[Doebner-Goldin  Equation for Electrodynamic Particle \& Applications/ JXZJ]{{\scriptsize{{Presentation at the Seventh Int. Conf. on Symmetry in Nonlinear Mathematical Physics, 2007}}} \\ \vspace{-0.6cm} \hrulefill\\  \vspace{0.8cm}

\title[Doebner-Goldin  Equation for Electrodynamic Particle \& Applications/ JXZJ]{Doebner-Goldin  Equation for Electrodynamic Particle. The Implied Applications$^*$ \\
{\small{With  Appendix: Dirac Equation for Electrodynamic Particles$^{**}$}}
}
\author{J.X. Zheng-Johansson$^{\dag, \ddag}$ }
\address{$^\dag$. Institute of Fundamental Physics Research,  611 93 Nyk\"oping, Sweden. \\ $^\ddag$. In affiliation with the Swedish Institute of  Space Physics, Kiruna, Sweden }
\address{September 30, 2007; updated January 28-29, 2008}

% In the case of the same organization, please use the following standard
%\Author{First Names LASTNAME and Second COAUTHOR}
%\AuthoqNameForHeading{F.N. Lastname and S. Coauthor}
%\Address{Address of Author(s), Country}
%\Email{email1@address, email2@address}
%\URLaddress{URL1, URL2)

%\ArticleDates{Received November 1, 2007, in final form  ????; Published online ????}

\begin{abstract}
We set up the Maxwell's equations and subsequently the classical wave equations for the electromagnetic waves which together with the generating source, a traveling oscillatory charge of zero rest mass, comprise a particle traveling in the force field of an usual conservative potential and an additional frictional force $f$; these further lead to a classical wave equation for the total wave of the particle. At the de Broglie wavelength scale and in the classic-velocity limit, the equation decomposes into a component equation describing the particle kinetic motion, which for $f=0$ identifies with the usual linear Schr\"odinger equation as we showed previously. The $f$-dependent probability density presents generally an observable diffusion current of a real diffusion constant; this and the particle's usual quantum diffusion current as a whole are under adiabatic condition conserved  and obey the Fokker-Planck equation. The corresponding  extra, $f$-dependent term in the Hamiltonian operator identifies  with that obtained by H.-D. Doebner and G.A. Goldin. The friction produces to  the particle's wave amplitude  a damping that can describe well the effect due to a radiation (de)polarization field, which is always by-produced by the particle's oscillatory charge in a (nonpolar) dielectric medium; such a friction and  the resulting observable diffusion as intrinsically accompanying the particle motion was strikingly conjectured  in the Doebner and Goldin original discussion. The radiation depolarization field in a dielectric vacuum has  two separate significances: it participates to exert on another particle an attractive, depolarization radiation force which resembles in overall respects Newton's universal gravity as we showed earlier, and it exerts on the particle itself an attractive, self depolarization radiation force whose time rate gives directly the frictional force $f$. 
\\ 
$^{*}, ^{**}$ Submitted to journal for publication. 
\\
$^{**}$ Included temporarily as Appendix I, pp. 20--36; to be submitted as a separate paper in near future.   
 \end{abstract}

           %\Keywords{Doebner-Goldin nonlinear Schr\"odinger equation; electrodynamic internal processes of particle; Maxwell's equations;  classical wave equation; electromagnetic waves; probability density current;   observable diffusion; adiabatic condition; Fokker-Planck equation;   dielectric (vacuum) medium; friction; radiation depolarization field;  self depolarization radiation force}
            %Please type here List of Keywords for your article separated by semicolon.

%\Classification{81P99;22E70} % e.g. 35A30; 81Q05
% For 2000 Mathematics Subject Classification see http://www.ams.org/msc/

%\maketitle

\def\Ical{{\mathcal{Z}}}

\def\Kcal{{\mathcal{K}}}
\def\kappab{\pmb{\kappa}}
\def\gd{{\mathcal{G}}}

\def\lb{{\bf l}}
\def\vb{{\bf v}}

\def\Rb{{\bf R}}
\def\pd{\partial}
\def\vphi{\varphi}
\def\psitot{{\mathcal{Y}}}

\def\psiR{\widetilde{\psi}}
\def\psiL{\widetilde{\psi}^{{\rm vir}}}
\def\psitotR{\widetilde{\psitot}}
\def\psitotL{\widetilde{\psitot}^{{\rm vir}}}

\def\PhimR{\widetilde{ {\mit \Phi}}}
\def\PsimR{\widetilde{ {\mit \Psi}}}
\def\PsimL{{\widetilde{ {\mit \Psi}}}^{{\rm vir}}}
\def\a{\alpha}
\def\uav{\bar{u}}
\def\D{\Delta}
\def\th{\theta}
\def\r{{\mbox{\tiny${R}$}}}
\def\re{{\mbox{\tiny${R}$}}}
\def\Fmed{F_{{\rm a.med}}}
\def\med{{\rm med}}
\def\Lw{L_{\varphi}}

\def\Efb{{\bf E}}
\def\Bfb{{\bf B}}
\def\Ac{ \varphi}
\def\Xsub{{\mbox{\tiny${X}$}}}
\def\Xssub{{\mbox{\tiny${X}$}}}
\def\Tssub{{\mbox{\tiny${T}$}}}
\def\Kb{{\bf{K}}}
\def\kb{{\bf{k}}}
\def\Ksub{{\mbox{\tiny${K}$}}}
\def\W{{\mit \Omega}}
\def\Wd{\W_d{}}
\def\Nu{{\cal V}}
\def\Nud{\Nu_d{}}
\def\Eng{{\cal E}}
\def\eng{{\varepsilon}}
\def\Acuni{\Ac_{{\Ksub}^\dagsup}^{\dagsup}}
\def\unduni{\Ac_{{\Ksub}^\dagger}^{\dagsup}}
\def\Acauni{\Ac_{{\Ksub}^\ddagsup}^{\ddagsup}}
\def\Acunim{{\Ac_{{\Ksub}^\dagsup}^{\dagsup *}}}
\def\undunim{{\Ac_{{\Ksub}^\dagsup}^{\dagsup}}^*}
\def\Acaunim{{\Ac_{{\Ksub}^\ddagsup}^{\ddagsup *}}}
\def\pd{\partial}
\def\Ad{ {\mit \psi}}
\def\psim{ {\mit \psi}}
\def\Kd{K_d{}}
\def\Lam{{\mit \Lambda}}
\def\lam{\lambda}
\def\dagsup{{\mbox{\tiny${\dagger}$}}}
\def\ddagsup{{\mbox{\tiny${\ddagger}$}}}
\def\psimKdK{\psim_{\Ksub,\Kdsub}}
\def\w{\omega{}}
\def\wdlow{\omega_d }
\def\g{\gamma{}} 
\def\Phim{{\mit \Phi}}
\def\Psim{{\mit \Psi}}     
\def\Psima{{\mit\Psi}}

\def\arm{{\rm a}}
\def\brm{{\rm b}}
\def\crm{{\rm c}}
\def\drm{{\rm d}}
\def\erm{{\rm e}}
\def\frm{{\rm f}}
\def\grm{{\rm g}}
\def\hrm{{\rm h}}
\def\lf{\left}
\def\rt{\right}
\def\Kdsub{{\mbox{\tiny${K_d}$}}}
\def\psimkd{\psim_{\kdsub}}
\def\psimKd{\psim_{\Kdsub}}
\def\hquad{ \ \ } 
\def\Taum{{\mit \Gamma}}

\section{Introduction}

While the usual linear Schr\"odinger equation has demonstrated to be adequate for the common nonrelativistic quantum  systems, L. de Broglie suggested \cite{dBroglie1960} in the 50s--60s that  the quantum mechanical  wave equation may be more generally nonlinear. Various forms of nonlinear equations have  been proposed and investigated subsequently, for a similar concern of  internal states of particle as L. de Broglie's or from rather different fundamental considerations. Of these, the Doebner-Goldin form of nonlinear Schr\"odinger equation,  Doebner-Goldin equation, represents a unique family  which  H.-D. Doebner and G.A. Goldin obtained in \cite{D-G} by 
admitting observable diffusion current to the probability density of a quantum particle, and subjecting this to  the continuity equation of a Fokker-Planck type on the basis of a unitary representation of an infinite-dimensional Lie algebra of vector fields and group of diffeomorphisms. 

In view that it admits observable diffusion which commonly  occurs to a greater or lesser degree in all macroscopic processes that are at the microscopic scale in majority cases executed by  quantum particles, and in view of the physical significance associated with the unitary representation of group theory 
 based on which it derives, the Doebner-Goldin equation can be anticipated to represent an important prediction of certain possible intrinsic processes accompanying quantum systems. What these possible processes  may be has on the other hand remained as  an open question prior to the present study. The main question may be formulated as that, what can be such an (intrinsic) dissipative process which disrupts not the stationary state of a quantum particle as an ordinary heat process would, 
 and which in the meantime manifests itself an observable diffusion? From a measurement point of view at least such a process is  viable even to a first degree, since such a process  would not cause any detectable effect if a  measurement is made over  the damped probability density current and the damping in amplitude does not change with time. A theoretical recognition of such processes however would seem unrealistic until recently, in view that the mechanical nature of the quantum processes described by the usual Schr\"odinger equation had remained up to interpretation. 

With overall experimental observations as input information the author recently proposed[{\citeUnif}a-e]   an internally electrodynamic (IED)  model for simple particles such as electron,   termed also basic particle formation (BPF) scheme in earlier reports [{\citeUnif}a-h, j]  (with coauthor P.-I. Johansson). 
The IED model, briefly, states that 
         %\begin{minipage}[t]{14.cm}
{\it a simple, single-charged particle  is constituted of an oscillatory point  charge $q$  of a zero rest mass  and the resulting electromagnetic waves propagating at the speed of light $c$}. 
          %\end{minipage}
In so far as the way the mechanism of the model operates, $q$ can be of arbitrary quantity;  $q$ is to be given as an input (out of  two sole input data, the other is the total energy of the charge)  to yield the actual material particles. 
[For examples, of the elementary  particles, for the isolatable charged one where the  multiple- or "neutral-" charged particles are viewed as achieved by   integration processes 
$n \leftarrow p+e+\nu_e$ and $N  \leftarrow p+n $, 
 clearly then $|q|= e$; for the, as of today, nonisolatable charged ones,  quarks, 
$|q|=1/3, 2/3$.] 
What form the basis of operating mechanism of the model merely  are three   elementary, experimentally firmly corroborated laws regarding electromagnetic waves: the Maxwell's equations 
           %or the derivative classical wave equation
 in respect of wave propagation,  the Doppler principle in respect of source motion effect, and the Planck energy equation in respect of energy discretization.

The obvious immediate motivation for proposing the IED particle model  was  to reconcile our understandings  of particles
with the various puzzling phenomena  involving particles  to date. For example it was not understood that what are the internal, mechanical processes which causes a particle to  manifest both as an extensive wave  and a point object depending on methods of detection? And what are such internal mechanical processes which appear to simultaneously also command a particle to emit or absorb electromagnetic waves through the exchange of a portion or the whole of its own internal energy or  inertial mass? What are the origin and nature of mass?
In view of the uniform presence of charges in all of material particles, of the universality of the vacuum as a medium to as far as we know, 
and of the fact that 
in ordinary connection to charge 
 the only pure waves 
 propagating in this vacuum  are electromagnetic waves, 
it is natural to expect  
that the internal processes of material particles are electromagnetic and that the operation of these hold the key to the answers to the various  relevant  puzzles.

In part as a broad  test of the IED model, and in part as an endeavor of the  understanding of a range of diverse phenomena from a common ground facilitated by it, 
a range of predictions of the 
fundamental  properties of particles and 
relations have been made in terms "first-principles" solutions for the IED partcile internal processes.  Here, the  "first-principles" refer to a minimal set of  firmly established physics laws consisting mainly the few aforementioned;  
 and the  charge and the total (mechanical) energy, ---corresponding to a characteristic oscillation frequency for a  universal vacuum medium, are as two sole input data.  
The achieved predictions [\citeUnif a-j]  include:  A particle has a spin and relativistic mass[\citeUnif d-e]  apart from the input characteristic charge and total energy;  
it is extensive as the result of its electromagnetic waves being extensive, and  when traveling freely  its waves evolve into a traveling, 
and in turn a standing, 
de Broglie phase wave between boundaries[\citeUnif c,e]. 
As such, 
the particle obeys the de Broglie relations[\citeUnif c,e];  
its  
traveling de Broglie phase wave will produce constructive 
interference at integer times the de Broglie wavelength upon superposition of its  different parts, 
and it in turn  behaves like a point object owing to its charge, say, when scattering elastically with another point particle.
            %striking at a detector into electric signal. 
More generally, in arbitrary potential fields under corresponding conditions the particle obeys the Schr\"odinger equation[\citeUnif d,e] and the Dirac equation[\citeUnif i]; 
the particle obeys the Einstein mass-energy relation[\citeUnif c,e], the Galilean-Lorentz transformation[\citeUnif f], and Newton's law of gravitation in attracting another particle[\citeUnif j],   among others.

 It is natural that we extend  in this paper  the studies   to aim to derive the  Doebner-Goldin nonlinear  Schr\"odinger equation, which we show will result when additionally subjecting the Schr\"odinger particle to a frictional force of the medium with the total system subjecting to an adiabatic condition. In part, this derivation provides an additional test of the IED basic-particle model. And in part, as the IED  model itself suggests  a natural origin of frictional force to be  the radiation depolarization field  always  produced in a dielectric medium 
       %side by side 
along with a particle's internal electromagnetic processes, the derivation and the solution will provide a formal elucidation for the  connection of this with 
          %the connection between this intrinsically accompanying process of an electrodynamical particle and 
the  Doebner-Goldin observable diffusion.

\section{Particle model with electrodynamical internal processes}\label{Sec-CEDPM}

We consider an IED model particle is traveling at a velocity $\vel$ as its oscillatory charge $q$ does,
for simplicity in a one-dimensional box along $X$-axis.  The oscillation of the charge is associated with a total energy $\eng_q$, 
$\eng_q$ being  smallest at $\vel=0$, $\eng_q |_{\vel=0}=\Eng_q$.
$\Eng_q$ or $\eng_q$ may be endowed e.g. in a pair production in the vacuum.   $\Eng_q$ 
 describes the ground state and 
           % by this virtue,  this energy 
therefore cannot be dissipated or detached from the charge except in a pair annihilation.

The charge will owing to its oscillation  generate electromagnetic waves, with the radiation electric field $\Eb^j$ and magnetic field $\Bfb^j$ (of the $j$ th component)  governed by the Maxwell's equations given in a medium of dielectric constant $\ke$ in  zero external fields as
$$ \displaylines{
\refstepcounter{equation} \label{eq-maxwel1}
\qquad \nablab \cdot \Eb^j =\frac{\rho_q^j}{\epsilon},\quad
 \nablab \cdot \Bb^j=0,
\quad
\nablab \times \Bb^j=\mu \jb_q^j  +\frac{1}{c^2}\frac{\pd \Eb^j}{\pd t},
\quad
\nablab \times \Eb^j =- \frac{\pd \Bb^j}{\pd t}.      
       \hfill (\ref{eq-maxwel1})
}$$  
Where $\rho_q^j$ is the density and  $\jb_q^j$ the current  of the  charge $q$ of the particle, assuming no other charges and currents present; $\epsilon=\ke\epsilon_0$ and $\mu\simeq \mu_0$, with $\epsilon_0$ and $\mu_0$ the permittivity  and permeability of the vacuum, 
         %represented in the usual way (see Sec. \ref{Sec-Apply} for alternative representations), 
and $c$ the velocity of light in the medium. 
          %%%%%%
          (Until the  specific application oriented 
 denotations 
 in Sec. \ref{Sec-Apply}, unless explicitly specified the 
respective media  for conveying the particle and for the reference of measurement will not be explicitly specified in writing the electromagnetic variables.)
              % resident medium for the particle and the reference medium for measurement are not and need not be explicitly specified; the dielectric and field variables all refer to this general context.) 
             %
Considering  only regions sufficiently away from the source charge so that $\rho_q^j=j_q^j=0$,  making some otherwise  standard algebra of the equations (\ref{eq-maxwel1}), and replacing the field variables by a more  general dimensionless displacement $\varphi^j$, with  $E^j=A\varphi^j $,   $B^j=E^j/c=A\varphi^j /c$ and   $A$ a conversion constant, we obtain the corresponding classical wave equation for each component electromagnetic wave $\varphi^j$:
$$ \displaylines{ 
\refstepcounter{equation}\label{eq-CMwave1}
\qquad \frac{\pd ^2 \varphi^j }{\pd T^2}=c^2 \nabla ^2 \varphi^j  \hfill (\ref{eq-CMwave1})
}$$
Here, in view of a Doppler effect to result from the source motion (to express explicitly later), we distinguish by the superscript $j$ the component wave generated in the direction parallel with the source velocity $\vel$,  denoted by $j=\dagger$, and the one in the direction antiparallel  with $\vel$, denoted by $j=\ddagger$;  within walls in stationary state there must also simultaneously prevail their reflected components and we may regard these as if being generated by a virtual charge (compared to the presently actual charge)
which is reflected and traveling in $-X$-direction, 
                   %(being virtual compared to the actual charge),  
denoting by  $j={\rm vir}\dagger$ and  $j={\rm vir}\ddagger$. Apparently, the total wave given by the sum of all of the component waves, $ \sum_j \varphi^j =\psitot_\subempty$,  describes the particle.

Based on the  general results of electrodynamics applied here to the particle's internal processes as governed by the basic  equations (\ref{eq-maxwel1})--(\ref{eq-CMwave1}),  basic properties of a given particle can be predicted; 
in the remainder of this section 
we outline two directly relevant ones of these. 
The first is  the total energy of the wave and accordingly the particle. As a general result of classical electrodynamics based on solution to the Maxwell's equations, the 
(\ref{eq-maxwel1}) here, combined with  Lorentz force law, an electromagnetic wave ($j$) transmits at the speed of light $c$ a  wave energy $\eng^{j}$ and  a linear momentum $p^j=\eng^j/c$.  In virtue of the stochastic  nature of the electromagnetic waves, its total  dynamical quantities, the total wave energy and linear momentum here, are appropriately the geometric means, 
$$\displaylines{\refstepcounter{equation}\label{eq-geomean}
\qquad
\eng=\sqrt{\eng^{\dagsup}\eng^{\ddagsup}}, \quad
p=\sqrt{p^{\dagsup}p^{\ddagsup}}; \quad
\sqrt{\eng^{\dagsup}\eng^{\ddagsup}}=\sqrt{p^{\dagsup}p^{\ddagsup}} c \quad {\rm or} \
\eng=p c.
\hfill (\ref{eq-geomean})
}$$ 
From the underlining laws afore-used, mathematically the amplitudes of $E^j, B^j$, $\varphi^j$, etc, and accordingly  $\eng^j$, $p^j$, $E$ and $p$  are permitted to take on continuous values. 

Following M. Planck's discovery of quantum theory in 1901, it has been additionally understood that the amplitudes of these quantities are  {\it in nature} quantized; the total wave energy of an electromagnetic wave of frequency $\w/2\pi$ is $\eng=n \hbar \w$, that is,  $\eng$  consists in general of $n$ momentum-space quanta, or photons, each of an energy $\hbar \w$.
The electromagnetic wave comprising our basic particle, like an electron, positron, etc., has, based on experimental indications  especially the pair processes,  
a "single energy quantum", $n=1$; the Planck energy equation for the total wave of the particle therefore is 
$$\displaylines{\refstepcounter{equation}\label{eq-engMP}
\qquad
\eng=\hbar \w.
 \hfill (\ref{eq-engMP})
}$$
  This total wave of a single energy quantum here
has  in a one-dimensional box  
two components, $\varphi^{\dagsup}$ and $\varphi^{\ddagsup}$, a situation no different 
          %obviously the same as 
from  discussed  after (\ref{eq-CMwave1}). 
Their wave frequencies are  Doppler displaced to  $\w^{\dagsup} $ and $\w^{\ddagsup} $  as a result of the source motion (to express explicitly below); and similarly as (\ref{eq-engMP}), 
 $\w^{\dagsup} =\eng^{\dagsup}/\hbar$, $\w^{\ddagsup} =\eng^{\ddagsup}/\hbar$. Further from (\ref{eq-geomean})  we have $\w=\sqrt{\w^{\dagsup}\w^{\ddagsup}}$. For the total wave comprising the particle, $\eng$ represents therefore  the total energy of the particle.  
It has been  proven and formally expressed especially through quantum electrodynamics that,  the Maxwell's equations  and naturally the derivative classical wave equation
continue to hold; 
and the quantization of the fields and the wave energy etc.  formally is the result of subjecting the corresponding canonical displacement and momentum, corresponding to the  $u(=a\varphi)$ 
 and $\dot{u}$ here, to the quantum commutation relation $[u,\dot{u}]=i\hbar$.  
In this generalized framework, clearly the classical solution of a continuous amplitude for say $\eng$ merely is an approximation when $n$ is large. 

The second property is the inertial mass of the wave and thus the particle. 
The two components of the electromagnetic wave, $(E^j,B^j)$ or $\varphi^j$,  rapidly oscillating at a geometric mean frequency $\w/2\pi$ and wavelength $\lam=c/(\w/2\pi)$,  viewed at some distance and ignoring the detail of the oscillations will appear as if being two  rigid objects, wavetrains, traveling at the speed of light $c$;  the two trains of the component waves together make a total wavetrain. In view that its speed of travel, $c$, is {\it finite} as contrasted to infinite, the total wavetrain has inevitably a {\it finite} inertia  mass, denoting this by $m$. This mechanical representation of the total wave, as a rigid "wavetrain", permits us at once to  express according to  Newtonian mechanics 
the linear momentum of the wavetrain to be $p=mc$. 
Combining this with the classical electrodynamics result  $\eng+0 =pc$ of (\ref{eq-geomean}) 
                 %(see Sec. \ref{Sec-CEDPM}.1), 
gives  the kinetic energy of the wavetrain $\eng =mc^2  $; this is just the Einstein's mass-energy relation (see e.g.   [{\RefUnif}c,g] for a detailed treatment).  This energy and the Planck energy of (\ref{eq-engMP})  ought to equal of course, thus  
$$\displaylines{
 \refstepcounter{equation}\label{eq-mhmu}
\qquad 
m= \hbar \w /c^2.
\hfill (\ref{eq-mhmu})
}$$
The  mass $m$ of the total wavetrain comprising the particle naturally represents the  mass of the particle, here  acquired dynamically through  the total motion of the waves of a geometric mean frequency $\w/2\pi$.  $m$ is 
dependent on the particle velocity and is thus 
relativistic, see further after equation (\ref{eq-K1}) below.

\section{
Wave equation of total motion of particle in external fields}\label{Sec.2}\label{Sec-eqomot}

To the  particle we now apply a Coulomb force $F=- \nabla V $ owing to a conservative scalar potential $V$,  and in addition a viscous force $f$;  
                %assuming  in the same direction, 
these give a total applied force  $F'=F+f$. We express the $f$ as follows. Suppose out of the total oscillation of the charge, a fractional displacement $u_q$ only produces radiation and is   defined here to be  equal to the wave displacement $u=a\psitot$, $\psitot $ being the dimensionless total wave displacement in the field of applied potentials; this in zero applied potential field is the $\psitot_\subempty$ earlier. Making direct analogy to the viscous force of ordinary mechanics, we can write down the frictional force opposing  the total motion of the particle as $f=\sum_n \frac{b_n}{(a\psitot)^n } (\frac{d (a\psitot) }{d T})^n$ in units of N. That is, in general $f$ is a function of the time rate of the total displacement  of the  charge or alternatively of the resulting wave displacement in the medium; section \ref{Sec-ViscForce} will give a concrete  representation of such a force. Assuming $d (a\psitot)/d T$ is small, so to a good approximation $ f=\frac{b_1}{\psitot \Lw} \frac{d \psitot }{d T}
=\frac{b_1}{\psitot \Lw} (\frac{\pd \psitot }{ \pd T}+ \nabla \psitot \frac{\pd X }{\pd T})$, where  $b_1$ is a constant in units of Nms and is real; $f$ is in units of $N$ and is generally imaginary (pointed out by D. Schuch)
for  $\psitot$ being generally  complex. This may rewrite as  
$$\displaylines{
\qquad f=-\frac{2mD}{\psitot \Lw}(\frac{\pd \psitot }{\pd T}+ \frac{\pd \psitot }{\pd X} \Vel)
\hfill
\cr 
{\rm where} \hfill
\cr \refstepcounter{equation}\label{eq-D1}
\qquad D=-\frac{b_1}{2m}, 
\hfill(\ref{eq-D1})
\cr 
\refstepcounter{equation}\label{eq-Vel2}
\qquad 
\Vel
=-\frac{ i \beta_1\hbar \vel_\obs}{2mD}-\Vel^*=\frac{ i\beta_1\hbar (\nabla \psitot^*) \psitot }{2m |\psitot|^2}. 
    \hfill (\ref{eq-Vel2})
}$$
$\Vel (\equiv \frac{\pd X}{\pd T}=\pd \w /\pd k)$ is the wave speed of $\psitot$, 
and $\Vel^*$ $(=-\pd X/\pd T=\pd \w' /\pd k')$ of that of the imaginary $\psitot^*
$. The expressions in (\ref{eq-Vel2}) follow firstly from the requirement that   $\Vel$ is in direct proportion with the velocity  $\vel_\obs$  of the current $j_\obs$ ($=\vel_{\obs} \rho$) of the  probability density $\rho$ ($=|\psitot|^2$)    in order to ensure the continuity of current  in a non-absorbing medium.  That is, $ \Vel = I \vel_{\obs}-\Vel^* $,  where the imaginary $\Vel^*$ is subtracted from the generally complex $I \vel_\obs$. The  current $j_{\obs}$ $=\vel_{\obs}\rho$ of  $\rho$ with a uniform translation  at velocity $\vel$  alternatively is  according to Fick's law the diffusion of a varying  $\rho$ in a viscous medium, $j_{\obs}=-D \nabla \rho$, with $D$ the diffusion constant. 
So,    
$\vel_\obs=\frac{j_\obs}{\rho}=-\frac{D \nabla \rho}{\rho} 
=-\frac{D}{\rho}[(\nabla \psitot^*)\psitot+\psitot^*\nabla \psitot]
$; 
here we have taken $D$  to be as defined in (\ref{eq-D1}) and this will receive a  justification later  through the role of $j_{\obs}$ in (\ref{eq-FP}). The above leads explicitly to the first and second expressions in (\ref{eq-Vel2}) once we  put the proportionality constant as  $I=-\frac{ i \beta_1\hbar }{2mD}$, where $\beta_1$ is a parameter yet to be determined  [by the equation (\ref{eq-FP}) below], and the other constants are  inserted so that  $\beta_1$ will have the  simple solution value 1.

In virtue of the electrodynamic origin of $F$  and inevitably also $f$ which are empirically established for point particles, extending  to the extensive IED particle here the two applied force and thus their  
total  $F'$ 
act apparently directly  on the point charge. 
We now want to map this  $F'$ into a force directly interacting with the $E^j,B^j$ or $\varphi^j$. We notice that  by its mathematical form equation (\ref{eq-CMwave1}) represents  just a classical wave equation 
for the electromagnetic wave  $\varphi^j$,  $a\varphi^j$ therefore
              % which accordingly represents  
a mechanical wave  propagated in an elastic medium, and  $F'$ interacts with $a\varphi^j$ by an  effective force $F'_{\med}$ acting on directly on this apparent medium. 
On the basis of this direct correspondence, but taken  as a heuristic means only in this paper (so that we need not to firstly introduce with sufficient justifications at any detail  the structure of this elastic medium), we shall below map the force $F'$ into $F'_\med$.
              %a force $F'_\med$ acting on this apparent   medium. 
Now, while $F'$ drives the charge, of a mass $m$ of the particle,  into an acceleration $\pd^2 \psitot/\pd T^2$, $F'_\med$ drives the medium of mass $\Mcal_\varphi$ (effectively) into acceleration $\pd^2 \psitot_\med/\pd T^2$ in $X$-direction.
Supposing the charge and the medium oscillate at a fixed phase difference if not in phase, the two  accelerations must be equal, we thus have
$$\displaylines{
 \refstepcounter{equation}\label{eq-fmed}\label{eq-Dp}
\qquad
F'_{\med} =\frac{\Mcal_\vphi}{m} (F+f)
=-\rho_l\lf[ \frac{V}{m} + \frac{ 2D}{\psitot } \lf(\frac{\pd \psitot }{\pd T }+ \Vel \nabla \psitot \rt)  \rt].
\hfill (\ref{eq-Dp})
}$$
Where $\Mcal_\varphi=\Lw \rho_l$, with $\rho_l$ the  linear mass density of the medium along the wave path of   a total  effective length $\Lw=J L $ ($\psitot$ winds in $J$ loops about the box side $L$).

We below further implement 
                      %need  further to translate
 the force $F'_\med$ in wave equation (\ref{eq-CMwave1})  similarly using the  heuristic approach by applying directly Newton's laws to the apparent elastic medium. In the apparent medium $\psitot$  corresponds to a physical, transverse displacement   $u=a\psitot=aC\sum_j \varphi^j$  as produced by the disturbance of the charge oscillation, with $a$ a conversion constant of length dimension. The deformed  elastic medium is consequently subject to  a tensile force $F_{\r}=\rho_l c^2$, with $c$ the velocity of light at which  $\varphi^j$ propagates. 
This force $F_{\r}$ and the applied force $F'_{\med}$ together give the total force acting on the particle through directly acting 
on the apparent elastic medium 
$$\displaylines{\refstepcounter{equation}\label{eq-Frp}
\qquad F'_{\r} = F_{\r} -F'_{\med}
= \rho_l \lf[
c^2 +\frac{ V}{m } 
+\frac{ 2D }{\psitot }\frac{\pd \psitot }{\pd T } \rt.$ 
$
\lf. + \frac{\beta_1Di\hbar }{m|\psitot|^2 }|\nabla \psitot|^2 
\rt], \hfill (\ref{eq-Frp})
}$$ where the minus sign in front of $F'_{\med}$ represents that this force tends to contract the chain. 
Consider on the linear chain of the medium  a segment $\D L$ at $(X,X+\Delta X)$  is upon deformation tilted from its equilibrium $\Delta X$ an angle $\Thm (X)$ and $\Thm+\D \Thm(X+\D X)$; 
assuming  $\psitot$ is small, $F_{\r}$ will be  uniform across the entire wave path $\Lw$.
The transverse ($Z$-) component force acting on it is 
$\Delta F'_{\r  t }=F'_{\r } [\sin(\Thm+\Delta \Thm)-\sin \Thm] 
$, 
with  
$
[\sin(\Thm+\Delta \Thm )-\sin \Thm ]  
= [1+O( \Thm)] \Delta \Thm  \simeq \Delta \Thm  
=\nabla^2 (a\psitot) \Delta R$;  $O(\Thm)$ collects the higher order terms and is dropped (this term leads to anharmonicity and not damping, and as  can be shown this in general leads not  to a Doebner-Goldin form of nonlinear term). Substituting in the above with (\ref{eq-Frp}) for $F'_{\r}$ we have    
$$\displaylines{\refstepcounter{equation}\label{eq-Frt}
\qquad
\Delta F'_{\r  t }
= F'_{\r }  \nabla^2 (a\psitot) 
\Delta R
=a\rho_l\lf[  c^2 
+\frac{ V}{m } 
+\frac{2D}{  \psitot}\frac{\pd \psitot }{\pd T } 
+ \frac{\beta_1Di\hbar }{m\rho }|\nabla \psitot|^2 
 \rt] \nabla^2 \psitot
\Delta R.
\hfill(\ref{eq-Frt})}$$ 
Applying Newton's second law  to the segment of a mass $\D \Mcal_\varphi = \rho_l \D L$, $\simeq \rho_l \D X$, 
we have  
$
\rho_l \Delta X \frac{\pd ^2 (a\psitot) }{\pd T^2}=\Delta F'_{\r t}$.
Placing  (\ref{eq-Frt}) in it, dividing $a\rho_l \Delta X$, we get  the equation of motion for per unit length per unit mass density of the elastic chain of medium at $X$, or equivalently the classical wave equation for the total (electromagnetic) wave of the particle:
$$\displaylines{\refstepcounter{equation}
\label{eq-eqmt1}
\qquad
\frac{\pd ^2 \psitot }{\pd T^2}    
=\lf[c^2 
+\frac{ V}{m }
+\frac{2D}{ \psitot }\frac{\pd \psitot }{\pd T } 
+ \frac{ \beta_1Di\hbar }{m\rho }|\nabla \psitot|^2 
\rt] \nabla^2 \psitot. 
        \hfill(\ref{eq-eqmt1})
}$$
In summary, (\ref{eq-eqmt1}) has a basic part 
$\frac{\pd ^2 \psitot_\subempty }{\pd T^2}    
=c^2  \nabla^2 \psitot_\subempty 
$ which one will get from summing  over all $j$ values the wave equations (\ref{eq-CMwave1}) given earlier directly from the Maxwell's equations in zero applied potential, and  it has an additional part describing the effect of the applied force $F'_\med$, derived with the help of the  "heuristic  elastic medium" approach.

Concerning the solution of (\ref{eq-eqmt1}), for the present we only consider explicitly the case of $D=0$. So (\ref{eq-eqmt1}) reduces to 
$\frac{\pd ^2 \psitot_\subempty }{\pd T^2}    
=\lf[c^2 
+\frac{ V}{m } 
\rt] \nabla^2  \psitot_\subempty$;  this being linear, thus $\psitot_\subempty=\sum_j\varphi^j$ and 
$$\displaylines{\refstepcounter{equation}\label{eq-vphi}
\qquad
                %\frac{\pd ^2 \psitot_\subempty }{\pd T^2}    =\lf[c^2 +\frac{ V}{m } \rt] \nabla^2  \psitot_\subempty, \quad {\rm or} \ \ 
\frac{\pd ^2 \varphi^j }{\pd T^2}    
                %%%%%%%%%%%%%%%
=\lf[c^2 
+\frac{ V}{m } 
\rt] \nabla^2  \varphi^j
\hfill (\ref{eq-vphi})
}$$
Supposing also $V$ is a constant, $V_c$,  equation (\ref{eq-vphi}) can be immediately solved to consist of  plane waves,  
$\vphi^{\dagsup}=C \exp[i({ k}^{\dagsup} X -\w^{\dagsup} T +\a_0)]$, $
\vphi^{\ddagsup}=-C \exp[i({k}^{\ddagsup} X +\w^{\ddagsup} T -\a_0)]$. 
Where $k^{j}=\g ^{j} K$ are the Doppler-displaced wavevectors for the wave generated parallel with the source velocity $\vel$
  ($j=\dagger$) and antiparallel with $\vel$ ($j= \ddagger$), and  $\w^{j} =\g ^{j} \W$ are the corresponding angular frequencies, with
$\g ^{\dagsup}=1/(1-\vel/c)$, $\g ^{\ddagsup}=1/(1+\vel/c)$. 
$K$ and $\W=Kc $ are the  values of $k^j$ and $\w^j$
at  $\vel=0$, $c$ being the velocity of light as before.
         %; $\W=cK$.   

The explicit superposition of the  incident waves $\varphi^\dagger,\varphi^\ddagger$ and their reflected ones ${\varphi^{{\rm{vir}}}}^\dagger,{\varphi^{{\rm{vir}}}}^\ddagger$ give a standing wave (for a systematic representation see  [\citeUnif a-c,e]):
 $$\displaylines{
\refstepcounter{equation}\label{eq-psitot}
\qquad \psitot_\subempty =\sum_j \varphi^j
=Ce^{i[(K+k_d) X]} e^{-i\w T},  
              \hfill (\ref{eq-psitot})
\cr
{\rm where} \hfill
\cr
\refstepcounter{equation} \label{eq-K1}
\qquad 
k_d=\sqrt{(k^{\dagsup}-K)(K-k^{\ddagsup})}=\g K_d, 
\quad K_d=\lf(\frac{\vel}{c}\rt) K;  
\quad \w=\sqrt{\w^{\dagsup}\w^{\ddagsup}}=\g \W; 
\hfill(\ref{eq-K1})
}$$
  $\g=\sqrt{\g^{\dagsup}\g^{\ddagsup}}=1/\sqrt{1-\vel^2/c^2}$. 
Canceling $\w$ between  (\ref{eq-K1}) and   (\ref{eq-mhmu}) gives further
$
                  %$\displaylines{ \refstepcounter{equation}\label{eq-mass1}\qquad
m=\g M  
$, 
with  $M= \hbar \W/c^2$  the classic-velocity limit ($\vel^2/c^2 \rightarrow 0$) of  $m$, i.e. the rest mass of the particle. 
An explicit inspection of (\ref{eq-psitot}) will  readily  show that $k_d$ is the de Broglie wavevector (for an existing elucidation see e.g. in [{\RefUnif}c,i]),  $K_d$ being its value at the limit $\vel^2/c^2 \rightarrow 0$.
        %For $V=V(X,T)$ being variant across $L$, the total wave function $\psitot$ can be represented as a Fourier sum of $\psitot|_{\Delta V(X,T)=V_c}$, to be expressed in  (\ref{eq-psimax2}) later. 
We shall later (see after equation \ref{eq-dgsch2}) generalize the representation to the case where $V$ may be arbitrarily varying in $L$; 
until then we shall proceed the following discussion for the constant $V$, $V_c$. 
Substitution of $\psitot_\subempty$ 
in the total wave equation given from the linear sum of the wave equations (\ref{eq-vphi}) over all $j$, i.e. with $D=0$ in (\ref{eq-eqmt1}),  directly  gives the expected relativistic energy-momentum relation for the particle[{\RefUnif}b], which  gives an additional check that  $\psitot_\subempty$ of (\ref{eq-psitot}) is the correct solution to the total wave equation.

For the solution of wave equation (\ref{eq-eqmt1}) with  $D$ finite we shall use the trial function:
$$ \displaylines{
\refstepcounter{equation}\label{eq-psi1} 
\qquad
\psitot=\Ical  \psitot_\subempty, \quad {\rm where } \quad
\Ical=e^{iQ}, 
\quad
Q=Q_1+iQ_2.  
\hfill(\ref{eq-psi1})
}$$
$\Ical$ represents a damping factor;  $Q_1$ and $  Q_2$ are real   variables and are in general functions of $X,T$. We shall restrict ourselves to the case where $D$ is small and accordingly $|iQ|<<|i(K+k_d)X-i\w T|$. Under such a condition, for the derivation of a nonlinear Schr\"odinger equation in question below, until the context of equation (\ref{eq-xxx2}),  an explicit solution form of $Q$ 
needs not be known.

\section{Transformation to wave equation for kinetic motion of particle}

At the classic-velocity limit $\vel^2/c^2 \rightarrow 0$, the total wave function $\psitot_\subempty$ reduces to (see [{\RefUnif}a-c])  $\lim_{\vel^2/c^2 \rightarrow 0}\psitot_\subempty=Ce^{i(K_d X-\Omegavel T)}$ with $\Omegavel (=\frac{1}{2}\W (\frac{\vel}{c})^2)=\frac{1}{2}K_d \vel+V_c $, which is  
equivalent to the solution for Schr\"odinger equation for an identical system as described by the wave equation (\ref{eq-eqmt1}) in the case  of $D=0$ and $V=V_c$.  
Therefore, as we noted in [{\RefUnif}a-c], 
equation  (\ref{eq-eqmt1}) must inevitably 
have a direct correspondence  
to the  Schr\"odinger equation,  and the remaining question mainly then was to identify a physically justifiable procedure to transform (\ref{eq-eqmt1}) to a form of the Schr\"odinger equation. 
Such a formal procedure  was elaborated in detail in [{\RefUnif}a-c]  
by a back-substitution of the explicit function $\psitot_\subempty$  in wave equation (\ref{eq-eqmt1}) in the case of $D=0$; by use of the Fourier theorem the procedure further led to a Schr\"odinger equation for 
 $V$  arbitrarily varying and also,  by a straightforward extension, for three dimensions.  
For the present case of $D$ being in general finite,
we below similarly first reduce and simplify  wave equation (\ref{eq-eqmt1}) at the classic-velocity limit $\vel^2/c^2 \rightarrow 0$ to a form such that the $K$- and $K_d$-
processes  are separable, by means of back-substitution of the formal function $\psitot$  (\ref{eq-psi1}) where      
       the function  $\psitot_\subempty$ is explicitly known  and  $Q$  assumed small.

We first prepare for the separation of the $K$- and $K_d$-processes in three aspects, the first two being similar as for the case $D=0$ [{\RefUnif}a,b]: 
(i) We observe that (\ref{eq-eqmt1}) contains the derivative 
$\frac{\pd^2 \psitot}{\pd T^2}$  which  relates to the acceleration of the particle and, as such, the $K$-  and $K_d$-  processes are not separable;  but the two processes are separable for the first derivative $\frac{\pd \psitot}{\pd T}$ which relates to the total energy  (for a detailed analysis see [{\RefUnif}b]). 
This suggests us to lower the time derivative one order as 
 $
\frac{\pd^2 \psitot }{\pd T^2}
\simeq \frac{\pd}{\pd T}
[(\frac{\pd \psitot_\subempty}{\pd T})e^{iQ}+\psitot_\subempty e^{iQ}i\frac{ \pd Q}{\pd T}]
\simeq  \frac{\pd}{\pd T}[-i\w \psitot+0] 
= -i\w \frac{\pd \psitot }{\pd T}
$.
(ii) In the two terms $\frac{ V}{m }\nabla^2 \psitot$ and 
$\frac{ \beta_1Di\hbar }{m\rho }|\nabla \psitot|^2 \nabla^2 \psitot $ in (\ref{eq-eqmt1}), the coefficients in front of 
$\nabla^2 \psitot$, being approximately the scale of quadratic thermal velocity $\vel^2$ or lesser,  
are relatively small for  $V$ and $D$ being small; and  also
                            %, as a weaker requirement, 
these are constant. 
So in these, consistent with the classic-velocity limit $\vel^2/c^2 \rightarrow 0$ in question, $\nabla^2 \psitot$ can to good approximation be replaced by its computed 
value:
 $
 \nabla^2 \psitot = \nabla[(\nabla \psitot_\subempty)e^{iQ}
+\psitot_\subempty e^{iQ}i\nabla Q]
\simeq 
 i(K+k_d)\nabla \psitot +0
= -(K^2 +k_d^2 )\psitot $, where
in going to  the second last expression we  dropped the cross-term products between the mutually orthogonal $\nabla e^{iKX} $ and $ \nabla e^{i k_d X}$  whose contribution to the final expectation value is in general zero (for an explicit proof see [{\RefUnif}a,b]). 
Using the identity relation $\g^2=1+\g^2 \frac{\vel^2}{c^2}$,  the above rewrites
$\nabla^2 \psitot
 =  -\g^2 K^2 \psitot  
$. 
(iii) 
In the two terms $\frac{2D}{ \psitot }\frac{\pd \psitot }{\pd T }  \nabla^2 \psitot$  and $c^2 \nabla^2 \psitot$ 
in (\ref{eq-eqmt1}), the coefficients $(\pd \psitot/\pd T)/\psitot \propto -i \w$ and  $c^2$ are large, with $\w$ being the scale of the particle's total energy. So,  
in these the $\nabla^2 \psitot$ 
ought to be kept  in functional form. 
But in the first of the two  terms  the large $\pd \psitot/\pd T$ itself  effectively will be unaffected by the small $V$ and $D$, and can therefore be  replaced  by its computed value $-i\w \psitot$ (used the small $Q$ assumption), thus 
$
\frac{2D}{\psitot}\frac{\pd \psitot }{\pd T}= -i2D\w
$. 
Substituting with the  reduced forms of (i)--(iii) 
for the respective 
$ \frac{\pd^2 \psitot }{\pd T^2}$, $ \nabla^2 \psitot$, and $\frac{2D}{\psitot}\frac{\pd \psitot }{\pd T}$ in (\ref{eq-eqmt1}), 
simplifying using the  basic relation $K^2\g^2 c^2 =\w^2$ and the relation   $mc^2=\hbar \w $ given in (\ref{eq-mhmu}),   
multiplying the resulting equation by
$-\frac{\hbar}{\w}$, (\ref{eq-eqmt1}) finally reduces to
$$\displaylines{
\refstepcounter{equation}\label{eq-tot1}
\qquad 
i \hbar \frac{\pd \psitot }{\pd T} 
=-\frac{\hbar^2 }{m}\nabla^2 \psitot
+V_c\psitot 
+i 2D\hbar  \nabla^2 \psitot 
+ i\beta_1D\hbar   \frac{ |\nabla \psitot|^2}{|\psitot|^2 } \psitot. 
               \hfill(\ref{eq-tot1})
}$$

We next proceed to separate in wave equation (\ref{eq-tot1}) the $K$- and the $K_d$- processes,  which are  inexplicitly contained in 
a  factor $\g$ in each term as we will  see explicitly below, and based on this we further reduce the equation at the classic-velocity limit. To this end, with $\psitot$ formally given in (\ref{eq-psi1}), we first compute each derivative in  (\ref{eq-tot1}) explicitly, and expand the $\g$ factor ($\g=1+\frac{1}{2}\frac{\vel^2}{c^2}+\frac{3}{8}\frac{\vel^4}{c^4} +\ldots$) in each:
$$\displaylines{\refstepcounter{equation}\label{eq-diffterms}
 \qquad
 \frac{\pd \psitot}{\pd T}
=-i \g \W \psitot 
= -i[\W+\Omegavel (1+\frac{3}{4}\frac{\vel^2}{c^2} +\ldots)]\psitot, 
\quad {\rm where }\ \ \Omegavel=\frac{1}{2}\W_d,\quad 
\W_d
         %=K_d \vel 
=\lf(\frac{\vel}{c}\rt)^2\W, 
\hfill
\cr
\qquad 
\frac{1}{m} \nabla^2  \psitot 
=-\frac{\g^2 K^2\psitot}{\g M}=-\frac{\g K^2\psitot}{ M}
=[-\frac{K^2}{M}-\frac{K_d^2}{2M}(1+ \frac{3}{4}\frac{\vel^2}{c^2}+\ldots) ]\psitot,
\hfill
\cr
\qquad
\nabla \psitot=i(K+\g K_d) \psitot, 
\quad
 |\nabla \psitot|^2
=(\nabla \psitot^*) (\nabla \psitot) 
=(K+\g K_d)^2 |\psitot|^2. 
\hfill (\ref{eq-diffterms})
}$$ 
On equal footing as the above,  $\psitot $ expands in its exponent as 
  $$\displaylines{
\refstepcounter{equation}\label{eq-Omegd}
\qquad 
\psitot   
= C e^{i [(K+\g K_d) X - (\W + \Omegavel(1 +\frac{3\vel^2}{4c^2}+\ldots) )T +Q] }.  \qquad
 \hfill (\ref{eq-Omegd})
}$$
The condition $\vel^2/c^2\rightarrow 0$  in general ensures 
$K>>K_d$, $\W>>>\W_d$. So, on the scales of $K_d$ and $\W_d$, the harmonic functions $e^{i K X}$ and $e^{-i \W T} $ oscillate so   rapidly that they present to any external observation effectively constants. Hence, 
 $e^{-i \W T} \simeq 1$,  $e^{i K X} \simeq1$; and  
 $$\displaylines{
\qquad
\refstepcounter{equation}\label{eq-pPsim}
\lim_{\vel^2/c^2 \rightarrow 0}\psitot   
= C e^{i [K_d X
                            -\Omegavel T +Q] }= \Ical \Psima_\subempty
\equiv \Psima,
 \quad 
\Psima_\subempty  = C e^{i [K_d X
                            -\Omegavel T] }.
            \hfill (\ref{eq-pPsim})
}$$
Taking accordingly the classic-velocity limit of the relations of (\ref{eq-diffterms}), substituting in the resulting relations with (\ref{eq-pPsim}) for $\Psim$ and its derivatives ($
 \nabla^2 \Psima =-K_d^2 \Psima$, 
  $\frac{\pd \Psima}{\pd T}
  = -i \Omegavel   \Psima$,   
$\nabla \Psima = i K_d \Psima$, 
$\nabla \Psima^* = -i K_d \Psima^*$, 
$|\nabla \Psima|^2 =  K_d^2 |\Psima|^2
$ for  small $Q$ assumption as earlier) for the $K_d$-, $\Omegavel$- terms 
while keeping  the $K$-,$\W$-terms as computed values which are large and will be unaffected for  $V$ and $D$ being assumed to be relatively small, 
 we have 
 $$\displaylines{
\refstepcounter{equation}\label{eq-exact0}
\qquad
    \lim_{\vel^2/c^2 \rightarrow 0}
 \frac{\pd \psitot}{\pd T}
              %%%%%%%%%
               %= -i \W   \lim_{\vel^2/c^2 \rightarrow 0} \psitot  -    \lim_{\vel^2/c^2 \rightarrow 0}i(\Omegavel+\ldots)\psitot = -i \W \Psima - i \Omegavel\Psima
         %%%%%%%%
=-i \W \Psima +\frac{\pd \Psima }{\pd T}, 
\ \  
   \lim_{\vel^2/c^2 \rightarrow 0}
\frac{ \nabla^2  \psitot }{m}
             %=-\frac{K^2  }{M}\lim_{\vel^2/c^2 \rightarrow 0}\psitot  -   \lim_{\vel^2/c^2 \rightarrow 0}\lf(\frac{K_d^2}{2M}+\ldots\rt) \psitot = -\frac{K^2\Psima}{M}  -\frac{K_d^2}{2M} \Psima\quad 
               %%%%%%%%%%%%% 
= -\frac{K^2\Psima}{M}  +\frac{\nabla ^2 \Psima }{2M},
\ \
\lim_{\vel^2/c^2 \rightarrow 0} 
        \nabla \psitot
               %%%%%%%%%
               %=iK\lim_{\vel^2/c^2 \rightarrow 0}\psitot +\lim_{\vel^2/c^2 \rightarrow 0} i\g K_d \psitot = i(K+ K_d) \Psima
               %%%%%%%
=iK\Psima +\nabla \Psima, 
\hfill
\cr
\qquad
\lim_{\vel^2/c^2 \rightarrow 0}
\nabla \psitot^*
        %%%%%%%%%
        %=-iK\lim_{\vel^2/c^2 \rightarrow 0}\psitot    -\lim_{\vel^2/c^2 \rightarrow 0} i\g K_d \psitot^*  = -i(K+ K_d) \Psima^*
        %%%%%%%%%
=-iK\Psima + \nabla \Psima^*,
 \quad
\lim_{\vel^2/c^2 \rightarrow 0} 
 |\nabla \psitot|^2
        %%%%%%%%
        %=K   \lim_{\vel^2/c^2 \rightarrow 0}  |\psitot|^2+ \lim_{\vel^2/c^2 \rightarrow 0}  \g K_d^2 |\psitot|^2= (K+ K_d)^2 |\Psima|^2?
         %%%%%%%%%
=K^2 |\Psima|^2+|\nabla \Psima|^2.
\hfill  (\ref{eq-exact0})
}$$
We dropped the cross-term products in the last relation of  (\ref{eq-exact0}) for similar consideration as earlier. Finally, subjecting wave equation (\ref{eq-tot1}) to the classic-velocity limit and substituting in the resulting equation with the expressions of (\ref{eq-exact0}) we have 
$$ \displaylines{\refstepcounter{equation}\label{eq-waveqtot}
\qquad
\hbar \W \Psima+i\hbar \frac{\pd \Psima}{\pd T}
= 
\frac{\hbar^2 K^2}{M} \Psima 
-\frac{\hbar^2 }{2M}  \nabla^2 \Psima
  +V_c \Psima     
-i 2D\hbar   K^2 \Psima 
+iD\hbar  \nabla^2 \Psima
\hfill
\cr
\qquad\hfill
+ \frac{   i\beta_1D\hbar      }{|\Psima|^2 }\lf[K^2\g^2 |\Psima|^2
+ 
{|\Psima|^2 } |\nabla\Psima|^2\rt] \Psima. 
\qquad\qquad (\ref{eq-waveqtot})
}$$
Equation (\ref{eq-waveqtot}) multiplied by $ \frac{1}{\Psima}$  contains a component equation
$$\displaylines{
\refstepcounter{equation}\label{eq-dgsch1}
\qquad  
\hbar \W  = 
\frac{\hbar^2 K^2}{M} 
    -i 2D\hbar   K^2  
+i\beta_1D\hbar  K^2
\hfill (\ref{eq-dgsch1})
}$$
 for a monochromatic electromagnetic wave produced by the given source but at zero velocity, and is not of our direct interest here. This equation holds always true for a given particle of a fixed rest mass and can be subtracted from 
$ \frac{1}{\Psima}\times$(\ref{eq-waveqtot}); multiplying $\Psima$ back to  the resulting equation from left, we obtain 
$$\displaylines{
\refstepcounter{equation}\label{eq-dgsch2}
\qquad
i\hbar \frac{\pd \Psima}{\pd T}= 
 -\frac{\hbar^2}{2M} \nabla^2\Psima
  +V_c \Psima     
+i D \hbar\nabla^2 \Psima
+ i\beta_1D\hbar   \frac{|\nabla\Psima|^2}{\rho } \Psima. 
\hfill (\ref{eq-dgsch2})
}$$

If $V$    varies arbitrarily  with $X$, thus $V=V(X,T)$,
$\vphi^{\dagsup}$ and $\vphi^{\ddagsup}$ are in general no longer plane waves. On the other hand, assuming $V(X,T)$ is well behaved, we can divide $L$ into  a large,  $N$ number of small divisions of width $\Delta X$ each. In each small division,  $(X_j,X_j+\Delta X)$, the potential, $V(X_j,T)=V_{cj}$, continues to be  approximately  constant and is  exactly so in the limit $\Delta X=  0$, and here the above plane wave method holds valid.  Elsewhere, $V(X_j,T)=0$. Going through therefore the foregoing procedure similarly for each  division, $j$, with  $j=1, \ldots, N$, we obtain equations of identical forms as (\ref{eq-pPsim}),  (\ref{eq-dgsch2}), etc., except with $\Psim$, $K_{d}$, $\Omegavel$ etc. denoted by $\Psim_{\Ksub_{dj}}$, $K_{dj}$, $\Omegavel_{\Ksub_{dj}}$,  etc. The $\{\Psim_{\Ksub_{dj}}(\Rb,T)\}$'s are mutually orthogonal and form a complete set. So the total wave function is the sum
$$\displaylines{
\refstepcounter{equation}\label{eq-psimax2}
\qquad
\Psim(X,T)
=\frac{1}{\sqrt{N}}\sum_{\Ksub_{dj}} 
A_{\Ksub_{dj}} \Psim_{\Ksub_{dj}}(X,T)
= \frac{1}{\sqrt{N}}\sum_{K_{dj}} 
A_{\Ksub_{dj}} C   e^{-i\Omegavel{}_{j} T +iQ} \cdot 
e^{i K_{dj} R};  \hfill
 (\ref{eq-psimax2})
\cr
{\rm or,} \hfill
\cr
\qquad \Psima=\Ical \Psima_\subempty, 
\quad
\Psima_\subempty=\Xima e^{-i\Omegavel T}, 
\quad
\Ical = e^{iQ_1 -Q_2},  
\quad
%\footnote{xxx}
\Xima
=  \frac{1}{\sqrt{N}}\sum_{K_{dj}} 
A_{\Ksub_{dj}} C 
e^{i K_{dj}  X-i(\Omegavel{}_{j}-\Omegavel) T},
\hfill
 (\ref{eq-psimax2})'
}$$ 
with $ A_{\Ksub_{dj}} Ce^{-i\Omegavel{}_{j} T +iQ}
=2\pi \sum_{s=1}^{N} \Psim(X_s,T) e^{-i K_{dj} \cdot X_s} $ 
 the Fourier transform of $\Psim(X_s,T)$. 
   %That a factor $e^{\Omegavel T}$ can be extracted out leaving the remaining time -independent in $\Xima$  of (\ref{eq-jim2}), is on the basis  that a stationary state particle has a well defined kinetic energy $\hbar \Omegavel $.

Multiplying $\frac{1}{\sqrt{N}}A_{\Ksub_{dj}}$ through the corresponding equation of (\ref{eq-dgsch2}) for $\Psim_j$,  
 summing the equations over all $j$ values we have
$$\displaylines{
\refstepcounter{equation}\label{eq-dgsch2b}
\qquad
i\hbar \frac{\pd \frac{1}{\sqrt{N}}\sum_j A_{\Ksub_{dj}} \Psima_j}{\pd T}= 
 -\frac{\hbar^2}{2M} \nabla^2 \frac{1}{\sqrt{N}}\sum_j A_{\Ksub_{dj}} \Psima_j
  +\sum_j V_{cj} \frac{1}{\sqrt{N}} \sum_j A_{\Ksub_{dj}} \Psima_{j}     \hfill
\cr
\hfill
+i D \hbar \nabla^2 \frac{1}{\sqrt{N}}\sum_j A_{\Ksub_{dj}}  \Psima_j
+ i\beta_1D \hbar  
 \frac{(\nabla \frac{1}{\sqrt{N}}\sum_j A_{\Ksub_{dj}}  \Psima_j)^*(
\nabla \frac{1}{\sqrt{N}}\sum_j A_{\Ksub_{dj}}  \Psima_j)
}{(\frac{1}{\sqrt{N}})^2\sum A^2_{\Ksub_{dj}} \rho_j } 
\frac{1}{\sqrt{N} }\sum_j A_{\Ksub_{dj}}  \Psima_j. 
\cr
\hfill (\ref{eq-dgsch2b})
}$$
Where,
$ \sum_j V_{cj}
=\ldots + 0\cdot V(X_{j-1},T) +1\cdot V(X_j,T) + 0\cdot V(X_{j+1},T) +\ldots
=V(X_j,T) 
$; $A_{\Ksub_{dj}}^*=A_{\Ksub_{dj}}$ since the amplitude of 
the physical displacement $\Psima$ must be  real; 
 $\rho=\sum_{j} \sum_j A^2_{\Ksub_{dj}} \rho_j=
\sum_j {A^*}_{\Ksub_{dj}} \Psima^*_j \sum_{j'}  A_{\Ksub_{dj'}} \Psima_{j'}  $ for $\Psima^*_j $ and $\Psima_{j'}$ mutually orthogonal and ${A^*}_{\Ksub_{dj}}$ real; and   
$\sum_j \sum_j |\nabla  A_{\Ksub_{dj}}  \Psima_j|^2
=
\nabla \sum_j (A_{\Ksub_{dj}}  \Psima_j)^*
\nabla \sum_j A_{\Ksub_{dj}}  \Psima_j $ for the two factors  mutually orthogonal. 

Substituting (\ref{eq-psimax2}) in  (\ref{eq-dgsch2b}) we obtain a generalized result of (\ref{eq-dgsch2}), a  wave equation  describing the  kinetic  motion of the particle
 in an arbitrarily varying, well-behaved potential $V$:
$$\displaylines{
\refstepcounter{equation}\label{eq-dgsch2bb}
\qquad
i\hbar \frac{\pd \Psima}{\pd T}= 
 -\frac{\hbar^2}{2M} \nabla^2\Psima
  +V \Psima     
+i D \hbar\nabla^2 \Psima
+ i\beta_1D\hbar   \frac{|\nabla\Psima|^2}{\rho } \Psima. 
\hfill  (\ref{eq-dgsch2bb})
}$$
Equation (\ref{eq-dgsch2bb})   is seen to represent an ordinary Schr\"odinger equation except for the  extra, nonlinear term 
$
%i\hbar G=
i D \hbar\nabla^2 \Psima
+ i\beta_1D\hbar   \frac{|\nabla\Psima|^2}{\rho } \Psima$ due directly to the frictional force  $f$.

\section{Diffusion currents. Continuity equation. Doebner-Goldin Equation}\label{Sec-ContEq}

Making some standard algebra to equation (\ref{eq-dgsch2bb}) and its complex counterpart   leads to an equation for the total current $j_{tot}=\Jcal_\qm+\Jobs$ of the probability density $\rho=|\Psima|^2$:
$$\displaylines{
\refstepcounter{equation}\label{eq-xx1}
\qquad 
\frac{\pd \rho}{ \pd T} +\nabla (j_\qm
+ \Jobs )
+(\beta_1-1 )2D \frac{|\nabla \Psima|^2}{\rho} |\Psima|^2=0.
\hfill(\ref{eq-xx1}) 
}$$
Where
$$\displaylines{
\refstepcounter{equation}\label{eq-curr}\label{eq-dens2}
\qquad 
j_\qm
=\frac{\hbar}{2Mi} [(\nabla \Psima^*)\Psima -\Psima^* \nabla \Psima], \quad
\Jobs=
-D \nabla \rho
=\frac{b_1}{2M}[(\nabla \Psima^*)\Psima+\Psima^*\nabla \Psima] 
\hfill (\ref{eq-curr})
}$$
with $j_\qm$  the usual quantum diffusion current 
and 
$\Jobs$ the observable diffusion current as earlier except now expressed in terms of the classic-velocity limit function   $\Psima$.
 The first quantity, $j_\qm$,  has an imaginary diffusion constant  $D_{\qm}=\frac{i\hbar}{2M}$ and this we know 
is to an external observer non-observable.

Suppose there are no "sinks" in the medium nor external reservoir in contact to the medium that trap or conduct the total current $j_{tot}$. So the particle and the (continuous) medium  as a whole is  adiabatic---a condition having an equal footing with the "unitary representation of vector field (of diffeomorphisms group)" employed in \cite{D-G}. Then, equation (\ref{eq-xx1})  for the total probability density current of the particle,  of a wave function $\Psima$ 
governed by wave equation (\ref{eq-dgsch2bb}), needs to conform to  the continuity equation which, for $D$ being real and $\Jobs$ being observable, is of the Fokker-Planck type: 
$$\displaylines{
\refstepcounter{equation}\label{eq-FP}
\qquad 
\frac{\partial \rho}{\partial T}  +\nabla (j_\qm
+ \Jobs )=0.
\hfill(\ref{eq-FP}) }$$
Comparison of this with (\ref{eq-FP}) suggests the third term in (\ref{eq-xx1}) must vanish; 
so  $\beta_1 = 1$.

With the $\beta_1$  value in  (\ref{eq-Vel2}) we find: 
$
\Vel+\Vel^*
=\frac{ -i \hbar \vel_\obs}{2mD}
             %=+\frac{ i\hbar\nabla \rho }{2m \rho} 
$,
$\Vel=+\frac{ i\hbar}{2m \rho}(\nabla \Psima^*) \Psima$, and 
$\Vel^*=+\frac{ i\hbar}{2m \rho}(\nabla \Psima) \Psima^*$. 
 With the $\beta_1$ value in turn directly in wave equation  
 (\ref{eq-dgsch2bb}), we finally have  
$$\displaylines{
\refstepcounter{equation}\label{eq-DG1}
\qquad 
i\hbar \frac{\pd \Psima }{\pd t}
=-\frac{\hbar^2}{2M}\nabla^2 \Psima + V\Psima +i D \hbar \nabla^2\Psima + iD \hbar (\frac{|\nabla \Psima|^2}{|\Psima|^2}) \Psima, \hfill (\ref{eq-DG1})
\cr
{\rm or} \hfill 
\cr
\qquad 
i\hbar \frac{\pd \Psima }{\pd t} = H' \Psima,
 \quad 
        %{\rm where} \ 
H'=H+iD\hbar G, \quad
H= -\frac{\hbar^2}{2M}\nabla^2+V, \quad G= \nabla^2\Psima+(\frac{|\nabla \Psima|^2}{|\Psima|^2}).
 \hfill
(\ref{eq-DG1}a)
}$$
Equation  (\ref{eq-DG1}) is seen to be exactly the Doebner-Goldin  form of nonlinear Schr\"odinger equation, the Doebner-Goldin equation, introduced in \cite{D-G}. In view of their respective meanings, the unitary representation of  vector fields in \cite{D-G} and the probability density conservation in an adiabatic total system here  are apparently two alternative but equivalent conditions. It is thus natural that the use of the latter here has led to the same result as based on the former  in \cite{D-G}. 

As was well appreciated in \cite{D-G} (1994), the nonlinear total Hamiltonian $H'$ as of (\ref{eq-DG1}a) is  in general complex and not Hermitian as would be required by the usual linear Schr\"odinger equation; a complex nonhermitian Hamiltonian is today a topic of increasingly many studies. 
               %see e.g. some recent publications by C.M. Bender relevant to the question whether such a Hamiltonian is CPT invariant. 
                 %%%%%%%%%%
In this regard the foregoing derivation of equation (\ref{eq-DG1}) based on the IED particle model additionally points to that, underlining the complex $H'$ and its imaginary part $iD\hbar G$ respectively are  a complex total force $F'_\med=F_\med +i|f_\med|$ and an  imaginary frictional force 
  $f_\med$, a property drawn the author's attention by D. Schuch at the SNMP conference, Kiev, 2007.  As is suggested by the mathematical form, we may comprehend the imaginary $f_\med$ as a physical variable  orthogonal to the real $F_\med$. As such, a measurement of the total force $F'_\med$ would then inform  the modulus  of it, $|F'_\med| =\sqrt{F^2_\med+f^2_\med}$, and not the direct addition of two scalar component forces and also not  an ordinary vector sum  ${\bf F}_\med' = {\bf F}_\med+ {\bf f}_\med$.

Concerning the solution for the Doebner-Goldin equation (\ref{eq-DG1}) we shall later only  refer to an interesting and also relevant case treated by  H.-D. Doebner and G.A. Goldin in \cite{D-G}. Starting with the denotations specified in (\ref{eq-psimax2})$'$ we have  the more general expressions: 
$\rho
= |\Xima|^2 e^{-2Q_2} $,
$\frac{\pd \rho}{\pd T}=-2|\Xima|^2 e^{-2Q_2}
 \frac{\pd Q_2}{\pd T}$,
$\Jobs=\frac{b_1}{2M}[(\nabla \Xima^* ) \Xima +\Xima^*\nabla \Xima
 +2  (\nabla Q_2)|\Xima|^2 ] e^{-2Q_2}
$,
$j_\qm
 = \frac{\hbar}{2Mi}[(\nabla \Xima^* ) \Xima -\Xima^*\nabla \Xima
 -2 i (\nabla Q_1)|\Xima|^2 ]e^{-2Q_2}
$.
Following\cite{D-G} we put  $Q_2=0$; the foregoing then  become: 
 $$\displaylines{
\refstepcounter{equation}\label{eq-jim1}
\refstepcounter{equation}\label{eq-jim2}
\qquad 
\rho= |\Xima|^2, \quad
\frac{\pd \rho}{\pd T}
=\frac{\pd |\Xima|^2}{\pd T}; \hfill (\ref{eq-jim1})
\cr
\qquad
j_\qm
 =\frac{\hbar}{2Mi}[(\nabla \Xima^* ) \Xima -\Xima^*\nabla \Xima
 -2 i (\nabla Q_1)|\Xima|^2 ], \quad 
\Jobs
=\frac{b_1}{2M}[(\nabla \Xima^* ) \Xima +\Xima^*\nabla \Xima
]. 
\hfill (\ref{eq-jim2}) 
}$$ 
Secondly,  suppose the system  is in stationary state, of which one of two possible descriptions  is $\frac{\pd \rho}{\pd T}=0$. (Since for $Q_2=0$, $\rho$ of (\ref{eq-jim1}) does not contain $D$ explicitly, so the "stationary state" here is not a small-$D$   approximation but  is exact as long as   (\ref{eq-DG1}) holds. But we derived  (\ref{eq-DG1}) based on a small $D$ condition, which agrees with the small $D$ requirement in \cite{D-G}.) Then  $\rho= |\Xima|^2 $ of (\ref{eq-jim1}) is  independent of time. Combining this with (\ref{eq-xx1})  follows $\nabla (j_\qm +\Jobs)=0$. 
Or, $j_\qm =D\nabla \rho  +B$ with $B$ a constant. 
 Substituting in  this last equation  with (\ref{eq-jim2}) for  
$ j_\qm$, restricting  $\nabla [(\nabla \Xima^*) \Xima -\Xima^* (\nabla \Xima)]=B$ to be  independent  of time which ensures  that when  $D= 0$,    $\Xima$ is a solution to the usual  Schr\"odinger equation, and further with the specific choice  $B=0$, one gets $\frac{-\hbar}{2Mi} (-2 i \rho \nabla Q_1)   = -D \nabla \rho$. Or,  
$\nabla Q_1=-\frac{\Taum \nabla \rho}{\rho}$ where $\Taum=mD/\hbar$. Integrating gives: 
$ 
Q_1 = -\Taum \ln |\Xima|^2
$. 
Substituting in (\ref{eq-psimax2})$'$
with this solution for $Q_1$ and the $Q_2=0$ earlier, one gets:
$$\displaylines{
\refstepcounter{equation}\label{eq-xxx2}
\qquad 
\Psima= \Ical \Psima_\subempty, \quad
\Psima_\subempty=
\xi e^{ -i\Omegavel T},
\quad \Ical =e^{-i\Taum \ln |\Xima|^2}.
\hfill(\ref{eq-xxx2})
}$$
Paper \cite{D-G} also discussed that the other of the two possible descriptions of the stationary state  to be $j_\qm=0$. This will also find a significant application in the examples later.  

Different forms of the nonlinear term would imply other boundary conditions than an adiabatic one, or other  applied forces than of the form here. In recent years, in terms of group theoretical approach H.-D. Doebner and G.A. Goldin \cite{D-G} (1994) and 
A.G. Nikitin and A.G. Nikitin and R.O. Popovych \cite{Nikitin2007} gave classifications of nonlinear Schr\"odinger equations in association with diffeomorphism group representations and in general terms.  
           %stemmed from a form of the nonlinear term in the current equation that can be symmetrically separated into the real wave and imaginary nonlinear Schr\"odinger equations, 
              %
D. Schuch gave an insightful review \cite{Schuch1994} on the nonlinear Schr\"odinger equations proposed by different authors with  
analysis regarding the quantum physical justifiability of solutions, 
                %raising especially the question of separability of the real and complex conjugate wave functions in the usually real and complex conjugate  Shr\"odinger equations and in connection a 
and  introduced an interesting logarithmic form of nonlinear Schr\"odinger equation.
H.-D. Doebner, A. Kopp and R. Zhdanov generalized in \cite{Nikitin2007}  nonlinearity to Dirac systems. 
Corresponding representations of these and beyond  
based on the IED particle model in the future  
can be similarly of value for test of the model and for gaining insight into the corresponding mechanical nature of  nonlinearity of quantum systems.

\section{Damping in dielectric media as 
 generic application of the Doebner-Goldin equation}
\label{Sec-Apply}

In most applications  the motion of a macroscopic object will in general be dissipated or more restrictively, damped, to a greater and lesser degree. The dissipation is typically known in the form of heat exchange with the environment and manifesting as an observable diffusion current.  But such a  description for a single quantum particle  system,   as described by (\ref{eq-DG1}) being in stationary state,  needs be taken in an effective, average way only. This directly follows from the circumstances that heat reflects  in general an energy current composed of many random collisions of a large population of individual (quantum) particles, during which the particles in general  deviate  from stationary state. Apart from its possible applications in the aforesaid effective way, it has been desirable\cite{D-G,D-GLect} to know  whether there may exist a Doebner-Goldin form of observable diffusion accompanying a stationary-state quantum particle literally and as an intrinsic process. The IED particle model underlining the foregoing derivation of the Doebner-Goldin equation implies in fact the presence of such processes applicable essentially to all quantum particles in any dielectric media;  we elucidate these below. 

Consider an IED particle is moving in the total  medium  of an ordinary material medium $n$ and the penetrating vacuum  taken here literally to be dielectric,  of a total dielectric constant $\ke$ as measured against  a true empty space, the space after removal of the dielectric vacuum. An explicit knowledge of the structure of the dielectric vacuum\footnote{
There exist today various propositions for the contents and structure of the vacuum  as held in different fields like in QED, QCD, etc., or by individual authors  including the "vacuuonic vacuum structure" proposed by the present author [\citeUnif g-h,e]; there  appears to exist no direct experimental information  regarding the explicit structure of the vacuum. 
}
is not needed  for the dielectric relations given in this paper.
The total medium and the particle are as a whole  evidently adiabatic. 
Measured against the true empty space, the particle's component radiation electric field propagated in the total dielectric medium is $E^\strempt$, and would be  $E_{\empty }^\strempt$ if  "propagated"\cite{Propagate} in the empty space. (In this section we shall for conciseness drop the superscript $j$, either because this is not directly of concern or the variables actually may represent the geometric mean quantities.) When  measured in the usual way against the vacuum with  the vacuum  regarded as effectively "non dielectric", the field $E$ in the total medium would be $E^\strvac$; $E^\strempt$ and  $E^\strvac $ represent the same force (note that the charge involved apparently causes no effect, for an original discussion see [\citeUnif1 e, g]) acting on the same medium as measured in the same inertial frame and must therefore be equal,  $E^\strempt \equiv E^\strvac$; this is  irrespective of against which medium the force is measured and represented. Supposing for simplicity the material medium is nonpolar,  with the vacuum being naturally nonpolar, so the total dielectric medium is nonpolar and the charge produces in it a depolarization field $E_{p}^\strempt$. The corresponding dimensionless  wave displacements accordingly are: $ \psitot^\strempt(\equiv \psitot^\strvac)=E^\strempt/A $, $\psitot_\empty^\strempt=E_\empty^\strempt/A $, 
and $ \Pw^\strempt=E_{p }^\strempt/A$.

Applying the standard dielectric theory for ordinary materials to the generalized dielectric system of an ordinary material and the vacuum here we can write down the following relations (for a systematic treatment see [\citeUnif g,e]): 
$ E^\strempt
(\equiv E^\strvac= \frac{E_{0}^\strvac}{\ke_n^\strvac })
= \frac{E_{\empty}^\strempt      }{ \ke^\strempt}$,
$
E_{p}^\strempt=-\chi^\strempt E$,
$
E=E_{\empty}^\strempt+ E_{p}^\strempt
$, with
$$\displaylines{
\refstepcounter{equation}\label{eq-dielrelc2}
\qquad
\ke^\strempt=\ke^\strvac_n  \ke_{_0}^\strempt,
\quad
\chi^\strempt
=\ke^\strempt -1
=(\chi^\strvac_n +1)(\chi_{_{0}}^\strempt+1)-1.
\hfill (\ref{eq-dielrelc2})
}$$ 
Where,  $\chi^\strempt$ is the susceptibility of the total dielectric medium and $\chi_0^\strempt$ that of the pure dielectric vacuum each measured against the empty space; $\ke_n^\strvac$ is  the dielectric constant and  $\chi^\strvac_n$ the susceptibility of the ordinary material medium $n$ measured in the usual way against a "non-dielectric" vacuum; $\ev_{_0}^\strempt $, $ (\equiv \ev_0^\strvac) =\ke_{0}^\strempt \ev_\empty^\strempt$, is the permittivity of vacuum and $\ev_\empty^\strempt$ the permittivity of the empty space. Multiplying  by  $1/A$, taking the classic-velocity limit as in (\ref{eq-Omegd}), the dielectric relations for the fields in the above become then 
$$\displaylines{
\refstepcounter{equation}\label{eq-dd1}
\qquad
\Psima
=\Psima_{\empty }^\strempt/\ke,
\quad
\Pw^\strempt=-\chi^\strempt \Psima, 
\quad
\Psima
           =\Psima_{\empty }^\strempt+ \Pw^{\strempt}.
\hfill  (\ref{eq-dd1})
}$$
Comparing the first relation in the above with 
the general form of wave function for the Doebner-Goldin equation, $\Psima=\Ical^\strempt \Psima_{\empty }^\strempt$  
of (\ref{eq-pPsim}) or more generally  (\ref{eq-psimax2})$'$,
we have 
$$\displaylines{\refstepcounter{equation}\label{eq-icalstar}
\qquad \Ical^\strempt=1/\ke^\strempt.
\hfill(\ref{eq-icalstar})
}$$
This states that, {\it the damping factor $\Ical^\strempt$  corresponds rather generally to the inverse of the dielectric constant $\ke^\strempt$  of the medium in which the particle resides.} In the case where $\rho^\strempt$ is independent of time, as specified by the  $Q_2=0$ and small $D^\strempt$ conditions,  $\Psima^\strempt$ and $\Pw^\strempt$ 
 are described  by the specific solutions (\ref{eq-xxx2}), which combined  with (\ref{eq-icalstar}) gives the corresponding expressions for the two dielectric parameters    
$$\displaylines{
\refstepcounter{equation}\label{eq-xxx1p}
\qquad
\ke^\strempt=1/\Ical^\strempt=e^{i\Taum \ln |\xi^\strempt |^2 },
\quad
\chi^\strempt =e^{i \Taum  \ln |\xi^\strempt |^2}-1.  
    \hfill (\ref{eq-xxx1p})                  
}$$ 

In the specific case when no ordinary  material presents, we have a pure dielectric vacuum,  thus $\ke_n^\strvac =1$, $\chi_{n}^\strvac =0$; (\ref{eq-dielrelc2}) and  (\ref{eq-dd1}) 
reduce to $\ke=\ke_0$, $\chi^\strempt =\chi_{0}^\strempt
=\ke_{0}^\strempt -1$,
              %$\Psima\rightarrow $ 
$\Psima_0=\frac{\Psima_{\empty}^\strempt }{\ke_{0}^\strempt}
=\Psima_{\empty }^\strempt+\Pw_{0}^\strempt $, 
            %$\Pw^\strempt\rightarrow $ 
$\Pw_{0}^\strempt=-\chi_{0}^\strempt \Psima_0$;
and  (\ref{eq-icalstar}) and  (\ref{eq-xxx1p})  reduce to 
$\Ical_0^\strempt=1/\ke_0^\strempt$.
$\Ical_0^\strempt$ and $\ke_0^\strempt$ are  for a specified particle here evidently universal constants, given that the vacuum is ubiquitous, isotropic and uniform throughout the space to as far as we know all of the time. As a consequence, the wave function $\Psima_{\empty }^\strempt$ appears to have never directly manifested itself in our present day's detections which are commonly based on the variation of the wave amplitude of a particle as a function of location and time; 
 our only direct knowledge of the particle wave appears to be the $\Psima_{0 }^\strempt$ ($\equiv \Psim^0_0$) of which the $\Ical_0$ or $\ke_0$ is  an inseparable component. Despite this, 
we see that first of all there presents a complete agreement between the prediction from the Doebner-Goldin equation, applied to the IED particle,  that the electromagnetic waves "inside" (or comprising) a particle can in general admit damping but without decaying with time in amplitude, and the fact that electromagnetic waves "outside" (i.e. detached from) a particle, becoming directly observable,  factually essentially do not decay with time in amplitude in the vacuum. 

Further, the dielectric vacuum, hence the $E_p^j$ field of a charge  in it and accordingly the $\Psima_{\empty }^\strempt$ wave in the empty space, has an indirect yet profound  manifestation according to a recent  theoretical prediction [\citeUnif j,f] by  the author  with coauthors. Namely,  the $E_p$ field participates to produce   an  attractive depolarization radiation force acted universally between two particles $1,2$ of masses $M_1$ and $M_2$ and charges $q_1$ and $q_2$, separated at a distance $R$. This force is as the result of the Lorentz force in their mutual    $E_{p}^\strempt$, $B_0$($\equiv B^\strvac$) fields say in the case of a pure vacuum: 
$F_{i i'}=q_j\frac{\Delta T q_{i'} E_{pi} B_{0i} }{M_{i'}}$, $ i,{i'}=1,2$.  
 The geometric mean of the mutual forces is 
$F_g =\sqrt{|<F_{12}><F_{21}>|}  =\frac{CM_1M_2}{ R^2}$, 
where  $< >$ represents  time average,  $|q_{i}|,|q_{i'}|=e$, 
 $C=\pi\chi_0 e^4/\epsilon_0^2 h^2 \rho_l $, 
 $e$ is the elementary charge and the other constants are as specified earlier; this  force $F_g$ was elucidated in  [\citeUnif j,f] to resemble in all respects Newton's universal gravity.  
To this application of the $E_p$ field, the present study  adds that the $E_{pi}{^j}^\strempt$ field of particle $i$ producing the depolarization radiation force leads directly to a  damping  
{\it in the particle's  Schr\"odinger wave $\Psim_{\empty i}^\strempt $}, by a factor  $\Ical^\strempt_0$, and the associated extra Hamiltonian term is a  Doebner-Goldin nonlinear term  added to that of the usual Schr\"odinger equation. In Sec. \ref{Sec-ViscForce} we shall  explicitly express the force directly responsible for the damping. Also in this context, the $j_\qm=0$ solution mentioned earlier may be a case where the particle wave and thus $j_\qm$ is shielded, say by  a material wall. And on the other side of the wall $j_\qm=0$; here  only the Doebner-Goldin observable diffusion current $j_\imr$ prevails. This directly corresponds to the property of the gravity which can not be shielded by any materials and on the other side of the wall as here it will propagate alone. 

In another specific case when an ordinary dielectric medium presents and we represent the vacuum in the usual way as non-dielectric which thus effectively plays the role of an empty space in the dielectric relations, the total dielectric medium thus reduces to the ordinary dielectric material medium alone, thus   
$\ke_0^\strempt=1$, $ \chi_0^\strempt =0$.  The relations of (\ref{eq-dielrelc2})--(\ref{eq-dd1})  now reduce to 
$\ke=\ke_n^\strvac$,
$
 \ke_n^\strvac -1 =\chi_n^\strvac$,
$\Psim_n=\Psim_0/\ke_n^\strvac$, 
etc., and (\ref{eq-icalstar}) reduces to 
$
1/\ke_n^\strvac =\Ical_n^\strvac
$.
Finally, with 
$
\rho^\strvac =|\xi^\strvac|^2
$, 
 (\ref{eq-xxx1p})  reduces to 
$\ke_n^\strvac =1/\Ical_n^\strvac =e^{i\Taum  \ln |\xi^\strvac |^2 }$, $\chi_n ^\strvac=e^{i \Taum  \ln |\xi ^\strvac|^2}-1$. 
That is, $\Ical_n^\strvac$ represents now a damping 
of the wave function $\Psima_0$ one would measure  in a pure vacuum medium, into $\Psim_n$ one will measure in the ordinary material medium of a dielectric constant $\ke_n^\strvac$. This is an usual representation of a particle in a material medium; this is directly comparable to the familiar fact  that a radiated (detached) electromagnetic wave in an optical material  in general experiences a complex dielectric constant and susceptibility. 

The above two specific situations would in general simultaneously enter in our material world, where a material particle commonly is more or less surrounded by other material particles or substances  and which, in the extreme case when all ordinary material substances are absent, is left to be the dielectric vacuum itself. Besides, a material particle is electromagnetic in nature (in the sense that such a particle  invariably  contains an electric charge),
               % and is described by the classical and quantum electrodynamic  theories in a unified framework), 
which is a direct observational fact and is irrespective of the specific IED particle model employed in this study (whereas the IED particle model only is essential in leading to a formal relationship between the  wave function, the electromagnetic field of the charge  and the corresponding depolarization field of  the particle).  
 These two universality features of the material world determine that a depolarization (radiation) field presents  intrinsically universally  with a particle. It therefore follows that a Doebner-Goldin damping, identified here with a depolarization radiation field, is an intrinsic phenomenon   presenting always to a material particle. Such a prospect that the nonlinear process could  be intrinsically universally accompanied with the particle process 
as elucidated in the above two examples  was strikingly conjectured in the  Doebner-Goldin original paper\cite{D-G}.

%\vfill  %**********************************************

\section{Self depolarization radiation force: Gravity from the  dielectric medium} \label{Sec-ViscForce}

We now give a concrete expression for the frictional force due to a total dielectric  medium, of dielectric constant $\ke$, against a particle moving in it. In the medium the particle's component radiation fields are $E^j, B^{ j}(=E^j/c)$,  $j=\dagger$ for the fields propagated in the direction parallel with $\vel$  and $j=\ddagger $ for fields antiparallel with $\vel$ similarly as earlier.  These fields are the results after  damping in the medium, from the un-damped  $E^j_\empty, B^j_\empty$ measured in empty space, by  a depolarization radiation field $E_{p}^{\strempt j}=-(E_\empty^{\strempt j}-E^j)=-\chi^\strempt E^j$ and a corresponding 
 $B_{p}^{^\strempt j}=-(B_{\empty}^{\strempt j}-B^j) = -\chi_{m }^\strempt B^j$ due to the presence of the medium,  
 where $\chi_{m}^\strempt= {\sqrt{\ke^\strempt }}^3 -1$, 
and $c=c_\empty/\sqrt{\ke^\strempt }$ (the permeability is assumed to be 1 here). The charge, due to   the $E_{p}^{\strempt j}$-,$B_{p}^{\strempt j}$- fields induced  by itself,  is   acted by a magnetic force 
according to the Lorentz force law: 
$$\displaylines{
\refstepcounter{equation}\label{eq-Fmp}
 \qquad 
\Fb_{m.p }^{ j}=q \velb_{p }^{\strempt j} \times \Bfb_{p }^{\strempt j} 
= {\rm (sign) }\frac{\chi^\strempt \chi_{m}^\strempt q^2 {E^j}^2}{Mc } \hat{X},
\hfill (\ref{eq-Fmp})
}$$  
where $\velb_p^\strempt=q E_{p}^\strempt /M$; sign$= +$ for $ j=\dagger$ and  $= - $ for  $ j=\ddagger$. Equation (\ref{eq-Fmp}) expresses that, irrespective of the  sign of the charge  and of the momentary directions of the alternating  fields generated by the charge to its right ($E^\dagsup,\pm B^\dagsup$ with a velocity $c$) and to its left ($E^\ddagsup,\mp B^\ddagsup$ with a velocity $-c$), $\Fb_{m.p }^{ j}$ is always a {\it pull} to the charge on either side from the medium. In this connection,   $\Fb_{m.p }$ refers to  a self depolarization radiation force  on the particle; and this  represents a gravitational force on the particle from the dielectric medium. 

While the particle is in stationary state, its (oscillatory) charge is constantly traveling in an alternating $+X$- and $-X$- directions.  The charge when traveling in the $-X$- direction  is similarly acted by a pull on either side, but now the ''sign'' $= +$ for $ j=\ddagger$ (wave generated opposite to the charge motion direction) and ''sign'' $= - $ for  $ j=\dagger$. So, on average the charge is acted  from either side by a pulling force, an attraction, given by the geometric mean of the two Doppler-displaced forces:
$$\displaylines{
\refstepcounter{equation}\label{eq-Fmpp}
\qquad
\Fb_{m.p }=\sqrt{\Fb_{m.p }^{ \dagsup} \Fb_{m.p }^{ \ddagsup}       }
= {\rm (sign)}\frac{\chi^\strempt  \chi_{m}^\strempt q^2 {E}^2}{Mc } \hat{X},
\hfill (\ref{eq-Fmpp})
}$$
where $E=\sqrt{E^{\dagsup}E^{\ddagsup}}$. This mapped to the medium is similarly  a pull to the deformed segment in question by its surrounding in the medium, corresponding to a reduced displacement of the medium,  this is opposite to the tensile force  associated in general with the usual $E^j$ field. 

We are here mainly interested in the resistance produced by  $F_{m.p}$  against the total motion of the particle, that is the time rate of $F_{m.p}$  as measured over a certain time interval $\Delta T$: $\int_0^{\Delta T} (d F_{m.p}/dT) d T\simeq \Delta T d F_{m.p}/dT$.  With this, the frictional force following the usual  definition is:
$$\displaylines{
\refstepcounter{equation}\label{eq-fmp}
\qquad
f_{m.p}=\frac{\Delta T}{\psitot}  \frac{d F_{m.p}}{d T}
                        %=\frac{1}{\psitot} \frac{d F_{m.p}}{d T}
 = \frac{\Delta T}{(E/A)} (
 \frac{\pd F_{m.p}}{\pd T}
+
\frac{\pd F_{m.p}}{ \pd X}\frac{\pd X}{\pd T}
)\hfill
\cr
\qquad \qquad
= \frac{\Delta T\chi^\strempt  \chi_{m}^\strempt q^2 A^2 }{Mc}
(
 \frac{\pd \psitot}{\pd T}
+\frac{\pd \psitot}{ \pd X}\Vel
) 
\hfill (\ref{eq-fmp})
}$$
where  $E=A \psitot$ and $\Vel=\pd X/\pd T$ as earlier. 
 Similarly as $F_{m.p}$,  $f_{m.p}$ is always  opposite in direction to the tensile force $F_\re$. We see that indeed,  in both its acting as a resistance against the particle total motion and in its functional form,   $f_{m.p}$ of (\ref{eq-fmp}) resembles fully the frictional force $f$ expressed formally in (\ref{eq-fmed}) earlier.

\subsection*{Acknowledgements}

The author would like to thank scientist P.-I. Johansson for his 
continued moral and funding support of the research, 
and  the Swedish Research Council for granting a Travel Grant 
that enables the author to travel to the 7th Int. Conf. on Symmetry in Nonlinear Mathematical Physics at Kiev for presenting this work and the Swedish Institute of Space Physics for administrating the grant. The author would like to thank  Professor D. Schuch for pointing out the frictional force is  imaginary, and for the subsequent   useful discussion regarding this and other related aspects 
                 % the imaginary force and the corresponding non-Hermitian Hamiltonian (a property which was clear from the beginning to HDD and also is closely related to some recent research aspects of CMB) etc. 
from  Professor D. Schuch,  Professor H.-D. Doebner, 
Professor Bender,   and Professor  Nikitin.   
The author would like to thank Professor H.-D. Doebner,   Professor J. Goldin
and Professor Dobrev for  valuable discussion and suggestions for a more informative and  adaptable  introduction to the particle model, and thank several distinguished Professors in Sweden for valuable reading of this and the related papers.

%\noindent{\footnotesize{ * The travel grant from the Swedish Research Council to the author is administrated by the Swedish Institute of  Space Physics, Kiruna, Sweden. }}

\vfill

%\eject
%\setcounter{equation}{0}
\setcounter{section}{0}

\hrulefill\\  
\section*{Appendix I}

\title[Dirac Equation for Electrodynamic Particles \quad /JXZJ]{Dirac Equation for Electrodynamic Particles}

\author{J.X. Zheng-Johansson
}
\address{
1.  
Institute of Fundamental Physics Research, 611 93 Nyk\"oping, Sweden;  
in affiliation with the Swedish Institute of  Space Physics, Kiruna, Sweden 
} 

\def\Ci{1}
\def\betamt{{\bf{b}}}
\def\kb{{\bar{k}}}
\def\kbf{{\bf{k}}}
\def\Kb{{\bf{K}}}
\def\cb{{\bf{c}}}

\def\pb{{\bar{p}}}
\def\pbf{{\bf{p}}}
\def\Acal{{\cal{A}}}
\def\Bcal{{\cal{B}}}
\def\Ccal{{\cal{C}}}
\def\Vp{V}
\def\m{{{}_{\mbox{-}}}}
\def\Ccal{{\cal{C}}}
\def\p{{{}_{+\hspace{-0.1cm}}}}

\def\psipi{\psi_{\p}(1)}
\def\psipii{\psi_{\p}(2)}
\def\psimi{\psi_{\m}(1)}
\def\psimii{\psi_{\m}(2)}

\def\ai{\alpha(1)}
\def\aii{\alpha^{'}(2)}
\def\bi{\beta^{'}(1)}
\def\bii{\beta(2)}

\def\fa{f_r}
\def\fb{f_\ell}

\def\Ca{C_a}
\def\Cb{C_b}
\def\fbf{{\bf{f}}}
\def\Ocal{{\cal{O}}}
\def\psib{{\pmb{\psi}}}
\def\alphab{{\pmb{\alpha}}}
\def\sigmab{{\pmb{\sigma}}}

\def\Eb{{\bf E}}
\def\Bb{{\bf B}}
\def\ke{\kappa}
\def\nabb{{\pmb{\nabla}}}
\def\nablab{{\pmb{\nabla}}}
\def\vir{{\rm vir}}
\def\psitot{\psi}
\def\jb{{\bf{j}}}
\def\vel{v}
\def\velb{{\bf{v}}}

\def\Imtr{I}
\def\citeUnif{2}
\def\App{}
\def\Qcal{{\mathcal{Q}}}
\def\Tcal{{\mathcal{T}}}
\def\Cross{Q}

\def\vphilim{f}
\def\ft{{\mathcal{B}}}
\def\vphibar{\mathbin{\varphi\mkern-12.5mu-}}

\def\vphi{\varphi}
\def\med{{\med}}

\def\Mcal{{\mathfrak{M}}}

\def\Sb{{\bf{S}}}
         \def\xia{{\mathcal{A}}}
\def\tha{\theta}

\def\nb{\bf{n}}
\def\zb{{\bf{z}}^0}
\def\phiv{\varphi}
\def\Lb{{\bf{L}}}
\def\velsub{_{\vel}}

\def\nablab{{\pmb{\nabla}}}
\def\velb{{\pmb{\vel}}}
\def\minus{\mbox{-}}

\def\Ab{{\bf{A}}_a}
\def\vel{\upsilon}
\def\Thm{\vartheta}
\def\lb{{\bf l}}
\def\vb{{\bf{v}}}

\def\Rb{{\bf R}}
\def\pd{\partial}
\def\vphi{\varphi}

\def\psitot{\varphi}
\def\psiR{\widetilde{\psi}}
\def\psiL{\widetilde{\psi}^{{\rm vir}}}
\def\PhimR{\widetilde{ {\mit \Phi}}}
\def\PsimR{\widetilde{ {\mit \Psi}}}
\def\PsimL{{\widetilde{ {\mit \Psi}}}^{{\rm vir}}}
\def\a{\alpha}
\def\uav{\bar{u}}
\def\D{\Delta}
\def\th{\theta}
\def\r{{\mbox{\tiny${R}$}}}
\def\re{{\mbox{\tiny${R}$}}}
\def\Fmed{F_{{\rm a.med}}}
\def\med{{\rm med}}
\def\Lw{L_{\varphi}}
\def\Fb{{\bf{F}}}

\def\Efb{{\bf{E}}}
\def\Bfb{{\bf{B}}}
\def\Ac{ \varphi}
\def\Xsub{{\mbox{\tiny${X}$}}}
\def\Ysub{{\mbox{\tiny${Y}$}}}
\def\Zsub{{\mbox{\tiny${Z}$}}}

\def\Ksub{{\mbox{\tiny${K}$}}}
\def\W{{\mit \Omega}}
\def\Wd{\W_d{}}
\def\Nu{{\cal V}}
\def\Nud{\Nu_d{}}
\def\Eng{{\cal E}}
\def\eng{{\varepsilon}}
\def\Acuni{\Ac_{{\Ksub}^\dagsup}^{\dagsup}}
\def\unduni{\Ac_{{\Ksub}^\dagger}^{\dagsup}}
\def\Acauni{\Ac_{{\Ksub}^\ddagsup}^{\ddagsup}}
\def\Acunim{{\Ac_{{\Ksub}^\dagsup}^{\dagsup *}}}
\def\undunim{{\Ac_{{\Ksub}^\dagsup}^{\dagsup}}^*}
\def\Acaunim{{\Ac_{{\Ksub}^\ddagsup}^{\ddagsup *}}}
\def\pd{\partial}
\def\Ad{ {\mit \psi}}
\def\psim{ {\mit \psi}}
\def\Kd{K_d{}}
\def\Lam{{\mit \Lambda}}
\def\lam{\lambda}
\def\dagsup{{\mbox{\tiny${\dagger}$}}}
\def\ddagsup{{\mbox{\tiny${\ddagger}$}}}
\def\psimKdK{\psim_{\Ksub,\Kdsub}}
\def\w{\omega{}}
\def\wdlow{\omega_d }
\def\g{\gamma{}} 
\def\Phim{{\mathcal C}}
\def\Psim{{\mit \Psi}}
\def\arm{{\rm a}}
\def\brm{{\rm b}}
\def\crm{{\rm c}}
\def\drm{{\rm d}}
\def\erm{{\rm e}}
\def\frm{{\rm f}}
\def\grm{{\rm g}}
\def\hrm{{\rm h}}
\def\lf{\left}
\def\rt{\right}
\def\Kdsub{{\mbox{\tiny${K_d}$}}}
\def\psimkd{\psim_{\kdsub}}
\def\psimKd{\psim_{\Kdsub}}
\def\hquad{ \ \ } 
\def\Taum{{\mit \Gamma}}

\begin{abstract}
We set up the Maxwell's equations and subsequently  the classical wave equations for the electromagnetic waves  which together with their  generating source, an oscillatory charge of zero rest mass in general travelling, make up a  particle travelling  similarly as the source at velocity $\vel$ in the field of an external scalar and vector potentials. The direct solutions in constant external field are Doppler-displaced plane waves propagating at the velocity of light $c$; at the de Broglie wavelength scale and expressed in terms of the dynamically equivalent and appropriate geometric mean wave variables, these render as functions  identical to the space-time functions of a corresponding Dirac spinor, and in turn to de Broglie phase waves  previously obtained from explicit superposition. For two spin-half particles of a common set of space-time functions constrained with antisymmetric spin functions  as follows the Pauli principle for same charges and as separately indirectly induced based on experiment for opposite charges, the complete wave functions are identical to the Dirac spinor. The back-substitution of the so explicitly  determined complete wave functions in the corresponding classical wave equations of the two particles, subjected further to reductions appropriate for the stationary-state  particle motion  and to rotation invariance when  in three dimensions, give a Dirac equation set; the procedure and conclusion are directly extendible to arbitrarily varying potentials by use of the Furious theorem and to three dimensions by virtue of the characteristics of  de Broglie particle motion.   Through  the derivation of the Dirac equation, the study hopes to 
lend insight into the connections between the Dirac wave functions and the electrodynamic components of simple particles under the government by the well established basic laws of electrodynamics. 

%\draft
%\keywords{ }

%\pacs{  %iop
%**     {PACS numbers:
        % { PACS numbers}:  
        % 41.20.Jb, %Electromagnetic wave propagation; radiowave propagation
        % 45.20.-d, %Formalisms in classical mechanics
        % 45.20.Dd, %Newtonian mechanics
        % 46.40.-f, %Vibrations and mechanical waves (under solids)
        % and 04.30.-w.  % Gravitational waves theory
        %40.00.00 ELECTROMAGNETISM, OPTICS, ACOUSTICS, HEAT TRANSFER, CLASSICAL MECHANICS, AND FLUID DYNAMICS 
 %**      03.50.De, %___Classical electromagnetism, Maxwell equations  (for applied classical electromagnetism, see 41.20.-q)
 %**      03.65.Ta, %___Foundations of quantum mechanics; measurement theory  (for optical tests of quantum theory, see 42.50.Xa) 
          %%%%03.65.-w Quantum mechanics  (see also 03.67.-a Quantum information; 05.30.-d Quantum statistical mechanics)
 %**       03.00.00, %___ Quantum mechanics, field theories, and special relativity  (see also section 11 General theory of fields and particles)
 %**       04.20.Cv, %___Fundamental problems and general formalism 
 %**       04.30.Db, %___Wave generation and sources 
%**       11.00.00, %___General theory of fields and particles  (see also 03.65.-w, Quantum mechanics and 03.70.+k Theory of quantized fields)
%**       41.60.-m, %___Radiation by moving charges
%**       41.20.Jb %___Electromagnetic wave propagation; radiowave propagation 
%**       }

\end{abstract}

%\submitted %iop

\section{Introduction}

P.A.M. Dirac established in \cite{Dirac1928A} a relativistic quantum mechanical wave equation, Dirac equation,  for a point electron based on the relativistic energy-momentum relation subjected to Lorentz transformation under rotation.
          %, or  invariance for small rotation of quantum theory. 
In \cite{Dirac1928A} P.A.M. Dirac  also theoretically predicted   for the electron the existence of an internal oscillation state,  a magnetic moment, and by interpretation of the negative energy solution, an anti-particle state known today as the positron. 
  The Dirac equation has proven to be  an accurate equation of motion  for (two) spin-half quantum particles at high velocities; most notably, Dirac predicted based on his equation  the relativistic intensities of Zeeman components of spectral lines and the frequency differences
              %\cite{Dirac1928A} 
[1 (1928b)] in exact  agreement with experiment. 
Up to the present however it has remained an open question  that what is waving with the Dirac wave functions, or Dirac spinor, a similar  question as for the Schr\"odinger wave functions and  the de Broglie waves ?
In addition, the Dirac theory meets with a few  its own open questions.
 What is the nature of a Dirac internal oscillation? 
             %How does the negative energy relate with the positive 
            %charge of positron?
            % What is (the nature of) a negative energy state 
            %if we hold that the electron hole picture is not fully 
            %  compatible with experimental observations ?
                   %%%%%%
How are the Dirac space-time functions explicitly connected with 
 the spin orientations, the signs of charges,  the signs of the energies, and in the extreme situation when an electron and position annihilate, the emitted two  gamma rays and conversely?
What is the  symmetry of the total spin of  an  electron and positron? 
Also, 
in the case of an isolated single electron or positron in zero external field where the spin orientation is of no consequence,  it would be desirable to have a way to directly write down the  corresponding Dirac equation without involving the Pauli matrices. These  as well as various other not fully addressed questions relating to fundamental physics seem to consistently point to the inadequacy of the point particle picture of today and  the need for a representation of the internal processes of the particles.

Recently, using overall experimental observations as input data  
the author proposed  an internally electrodynamic (IED) particle model [\citeUnif a] (with coauthor P.-I. Johansson) or  sometimes  termed a basic particle formation (BPF) scheme, which states that {\it a simple (basic) particle like an electron and positron, etc.,  briefly, is constituted of an oscillatory  point-like (elementary) charge with a specified sign and a zero rest mass, 
and the resulting electromagnetic waves in the vacuum.  }
           %The BPF scheme is supplemented by a model vacuum 
           %[{\citeUnif}c,e]that  is filled of neutral and polarizable 
           % entities, vacuuons. 
          %%%%%
As a broad test of the IED particle model and also as an endeavour of understanding the various puzzles relating to fundamental physics, in terms of solutions for the electrodynamic processes of the model particle with its charge's sign and  total energy as two sole input data, 
            the author has further achieved with coauthor(s) 
derivations/predictions of a range of basic properties  and relations of the simple particles [\citeUnif a-j] 
               %including the Schr\"odinger equation 
              %This is  along with the predictions/derivations of a range of other fundamental properties of the basic particles and relations achieved by the author with coauthor(s) [\citeUnif a-b, d-k] 
               %%%%%%%%%%%
including the 
relativistic mass, de Broglie wave, de Broglie relations, Schr\"odinger equation, 
Einstein energy-mass relation, Newton's law of gravity and Doebner-Goldin equation, among others.  
 As to the Schr\"odinger wave function specifically relevant here, the solution[\citeUnif a,c] showed that it is the (envelope of the) standing wave, superposed from the Doppler-differentiated electromagnetic waves generated by the particle's travelling source charge,  that is  waving.

As previously shown e.g. in [\citeUnif c], the direct solutions for the classical wave  equations, derivable from the Maxwell's equations, for the electromagnetic waves comprising a free particle consist of  Doppler-displaced plane waves; these   superpose to two opposite-travelling  beat waves that resemble directly the de Broglie phase waves and in turn the Dirac space-time  functions which in common are functions of the particle's total energy and linear momentum and thus are "relativistic". It therefore is foreseeable that  the classical wave equations for the electromagnetic waves would more naturally lead to a wave equation of the particle corresponding directly to the Dirac equation in comparison to the Schr\"odinger equation. We elucidate  in this paper a formal procedure which transforms the classical  wave equations for the electromagnetic waves of two spin-half particles, of identical space-time functions and tending to approach one another, to the Dirac equation. Through the procedure 
%especially
 we show that the Dirac internal oscillation  corresponds to the oscillation of the electromagnetic waves at a geometric mean of frequencies which in general are Doppler-displaced owing to source motion,  
              %and the internal electromagnetic waves,  charges and spins composing two  identical spin-half particles relate with one another according to the few very basic laws of electrodynamics.
              %%%%%%%%%%%%%%%%%%%%%
and we elucidate the explicit relationships between the internal electromagnetic waves, charges, 
 spins, the centre-of-mass and total wave motions and the associated energies  of the particles    under the government of a few established elementary laws of electrodynamics.

\section{Wave equations for the electromagnetic waves of particle. Solutions}
\label{Sec-IED-particle}

We consider an IED particle, here an electron or positron, is as  its source charge $q$ ($=e$ or $-e$)  travelling at a velocity $\velb$ in $+z$-direction for the present along  a one-dimensional box of side $L$ in the vacuum. The  charge $q$ of the particle has an oscillation  associated with a total energy $\eng_q$, which is minimum  at $\vel=0$, denoted by $\Eng_q$; $\Eng_q$ may be endowed e.g. in a pair production  in the vacuum. In virtue that it describes the ground state, $\Eng_q$ cannot be dissipated or detached from the charge except in a pair annihilation. 

The  charge $q$ of the particle  generates owing to its oscillation electromagnetic waves of radiation electric fields $\Eb^j$'s and magnetic fields $\Bfb^j$'s described in zero applied potential field by  the  
Maxwell's equations  as:
$$ \displaylines{
\refstepcounter{equation} \label{eq-maxwel1}
\hfill \nablab \cdot \Eb^j =\rho_q^j/\epsilon_0,\ 
 \nablab \cdot \Bb^j=0,
\ 
\nablab \times \Bb^j=\mu_0 \jb_q^j  +(1/c^2)\pd_t \Eb^j,
\ 
\nablab \times \Eb^j =- \pd_t \Bb^j.      
       \hfill (\ref{eq-maxwel1})
}$$  
Where $\rho_q^j$ is the  density and  $\jb_q^j$ the current  of the particle's charge, assuming no other charges and currents present; $\epsilon_0$ is the permittivity 
and $\mu_0$ the permeability of the vacuum, and $c$ is the velocity of light; $\pd_t \equiv \pd /\pd t$. Expressing the $j$th fields  generally by a dimensionless displacement $\varphi^j$, $E^j=D\varphi^j $, thus $B^j=E^j/c=D\varphi^j /c$, with $D$ a conversion constant, considering  regions sufficiently away from the source only  so that  $\rho_q^j=j_q^j=0$, and with some otherwise  standard algebra of the Maxwell's equations (\ref{eq-maxwel1}),  we obtain the corresponding classical wave equations  for the  electromagnetic waves $\varphi^j$'s
$$ \displaylines{ 
\refstepcounter{equation}\label{eq-CMwave1}
\hfill 
c^2 \nabla ^2 \varphi^j=\pd_t ^2 \varphi^j,  \hfill (\ref{eq-CMwave1})
}$$
with $\pd_t^2\equiv \pd^2/\pd t^2 $. In the above,  $j=\dagger$ labels the component wave generated in the direction parallel  with  $+\vel$, and $j=\ddagger$  the wave parallel with $-\vel$;  within walls there prevail  also their reflected components  generated by the reflected charge at an earlier time and being at the present time as  if generated by a virtual charge travelling in the $-z$-direction,  labelled by $j={\rm vir}\dagger$ and  $j={\rm vir}\ddagger$. $j$ is to distinguish a Doppler effect owing to the source motion to be expressed in  (\ref{eq-Doppler}) below. In \ref{Sec-wave-energy} we outline in relevance to the particle model a few further standard relations of classical and quantum electrodynamics for the electromagnetic waves, and a derivation of the particle's mass given by the author previously[\citeUnif a,e] (with P.-I. Johansson). 

To the particle we now apply an electromagnetic  force $\Fb=\Fb_e+\Fb_m$, with $\Fb_e=- q\nablab \phi_a $ the Coulomb force in $z$- direction and  $\Fb_m=-q \vb \times \nablab \times \Ab $  the Lorentz force  due to an external scalar potential $\phi_a$ and vector potential $\Ab$, expressed in SI units as for all other quantities in this paper. $\Fb_m$ may be simplified using the BAC-CAB rule as $\Fb_m=-q[\nablab (\velb \cdot \Ab)-\Ab (\velb \cdot \nablab)]$.  In the applications below $\Ab$ is constant 
in $L$ or in each  small division in question (see end of Sec. \ref{Sec-Mean}), and $\vel$ is constant in $L$ for the particle being in stationary state and  also is parallel with $\nablab$ and ${\bf z}$, so $\nablab (\velb \cdot \Ab)=0$ and $\Fb_m=q\Ab (\velb \cdot \nablab)=q\Ab \vel \nabla
$. Thus, $\Fb=- q\nablab \phi_a+q\Ab \vel \nabla
$. 
The formula of $\Fb$ is in the usual usage established for a point particle; so when extending to the extensive IED  particle here, $\Fb$ apparently directly acts  on the particle's point charge.

We need to map the $\Fb$ to a force directly interacting with the internal fields  $E^j,B^j$, or $\varphi^j$ of the particle.    We observe that, in virtue of its form, (\ref{eq-CMwave1}) represents  just a classical wave equation for a mechanical wave of a transverse displacement $a\varphi^j$ propagated in an apparent {\it elastic medium}, $a$ being a conversion factor of length dimension and apparently being cancelled in (\ref{eq-CMwave1}). On grounds of this direct correspondence, but taken as a heuristic means only in this paper (so that we here need not involve the details of this elastic medium), $\Fb$ therefore interacts with the internal fields through a force $\Fb^j_\med$ directly acting on this  apparent medium. We can think of the medium to be composed of coupled dipole charges which do not move along the $z$-axis but the $\Fb^j_\med $ propagates across the dipoles at the wave speed $c$. If viewing  in a frame where $\Fb^j_\med $ is at rest, then effectively the dipole charges are travelling at the speed $c$; 
so as a first step of mapping, the Lorentz force on the medium ought to scale  as $\Fb{^j}'_m=(\pm c/\vel)\Fb_m
=\pm q\Ab c \nabla$ with $+,-$ for the $j=\dagger,\ddagger$ waves; thus $\Fb{^j}' = \Fb_e+\Fb{^j}_m'$.
Under the actions of the respective forces, the acceleration $F{^j}'/m^j$ of the particle's charge of a dynamical mass $m^j$   (due to the charge's total motion and equivalently  the $\varphi^j$ motion, see further \ref{Sec-wave-energy}), and that of the medium of a dynamical mass $\Mcal_\varphi^j$, $F^j_\med/\Mcal_\varphi^j$ must equal, i.e. $\Fb{^j}'/m^j=\Fb^j_\med/\Mcal_\varphi^j$. Thus   
$\Fb_{\med}^j = \frac{\Mcal_\vphi^j}{m^j}\Fb{^j}'
=\frac{\Mcal_\vphi^j q}{m^j \Lw^j} (-\nablab \phi_a  \pm\Ab c \nabla)$.
 
By its pure mechanical virtue the force $\Fb_{\med}^j$ acting on the continuous medium is nonlocal and will be transmitted uniformly  across the  medium here along the $z$-axis of effective lengths $\Lw^{\dagsup},\Lw^{\ddagsup}$ for  the $j=\dagger,\ddagger$ waves ($\psitot^j$ winds $J^j$ loops about $L$). 
Using the geometric mean $\Lw=\sqrt{\Lw^{\dagsup}\Lw^{\ddagsup}}$,  thus  $\nablab \phi_a= \pm (\phi_a /\Lw ) \hat{z}$ and  $\nabla =\pm 1 /\Lw$. With these,  putting  $\Mcal_\vphi^j=\rho_{_{l}} \Lw^j$ where $\rho_{_{l}}$ is the (geometric mean) linear mass density of the medium, writing for conciseness $\nablab \phi_a$  and also the final $\Fb_{\med}^j$  in  scalar forms and keeping the generally arbitrarily oriented $\Ab$  in vector form only, $\Fb_{\med}^j$ becomes
$$\displaylines{\refstepcounter{equation} \label{eq-x1}
\hfill
F_{\med}^j 
=-\rho_{_{l}} \Vp^j/m^j, \quad 
V^\dagsup=q\phi_a - q\Ab c, \quad 
V^\ddagsup=-q\phi_a -q \Ab c.
\hfill (\ref{eq-x1})
}$$

 $F_\med^j$ can be implemented in (\ref{eq-CMwave1}) by directly establishing the corresponding wave equation for the apparent elastic medium acted by $F_\med^j$. If without $F_\med^j$, 
the   elastic medium would be deformed 
owing to the disturbance of the oscillation of the source charge alone, 
by a total displacement $u=a\sum_j \varphi^j_{\vel'}$,
                % with $C$ a normalisation factor, 
and be thus subject to  a tensile force $F_{\r}=\rho_{_{l}} c^2$. 
The applied $F_\med^j$ and $F_{\r}$ add up to a total force acting on the particle through acting directly on the medium
$$\displaylines{\refstepcounter{equation} \label{eq-x2}
\hfill
F^j_{\r}{}' = F_{\r} -F^j_{\med} 
= \rho_{_{l}} \lf[
c^2 +\Vp^j/m^j \rt].
\hfill (\ref{eq-x2})
}$$
Where, the minus sign of $F_{\med}^j$ is because this force 
tends to contract the chain.
 Assuming $\psitot^j$ is relatively small which in general  is the case in practical applications,  $F^j_{\r}{}'$ is thus uniform across the $L$. 
A segment $\D L$ of the medium along the box, of mass  $\D \Mcal_\vphi =\D \Mcal_\vphi^j /J^j= \rho_{_{l}} \D L \simeq \rho_{_{l}} \D z$, will upon deformation be tilted from its equilibrium  position $z$-axis  an angle $\Thm^j $ and $\Thm^j+\D \Thm^j$  at $z$ and $z+\D z$. The transverse ($y$-) component force acting on $\D \Mcal_\vphi  $ is 
$
\Delta F^j{}'_{\r  t }=F^j{}'_{\r } [\sin(\Thm^j+\Delta \Thm^j)-\sin \Thm^j] 
= F^j{}'_{\r }  \nabla^2 (a\psitot^j) 
\Delta z
=a\rho_{_{l}}\lf[  c^2 
+\frac{\Vp^j}{m^j} 
 \rt] \nabla^2 \psitot^j
\Delta z
$. 
 Newton's second law for the mass $\D \Mcal_\varphi$ writes 
$
\rho_{_{l}} \Delta z \pd_t ^2 (a\psitot^j) =\Delta F^j{}'_{\r t}$. The two last equations give the equations of motion, on dividing $a\rho_{_{l}} \Delta z$,
for per unit length per unit linear mass density of the medium  at $z$ or equivalently the classical wave equations 
for the electromagnetic waves 
$\varphi^{j}$'s in the fields of the applied potentials $\phi_a,\Ab$:  
$$\displaylines{\refstepcounter{equation} \label{eq-eqm1}
\hfill
\lf[c^2+q(\phi_a - \Ab c)/m^{\dagsup}  \rt]
\nabla^2 \varphi^{\dagsup} 
             =\pd^2_t \varphi^{\dagsup}, 
\quad
[c^2 -q(\phi_a + \Ab c) /m^{\ddagsup}]
\nabla^2  \varphi^{\ddagsup} 
=  \pd_t^2 \varphi^{\ddagsup}.  
\hfill (\ref{eq-eqm1})
}$$
This for $\phi_a=A_a=0$ reduces to (\ref{eq-CMwave1}) given directly from the Maxwell's equations earlier.  

Assuming for the present  $\phi_a, \Ab$ are constant and also  $\Ab$ is  small such that the particle motion effectively deviates not  from the linear path,
so the solution of (\ref{eq-eqm1}) consists of plane waves 
$\varphi^{\dagsup}=\Ccal f^{\dagsup}$ 
and  
$\varphi^{\ddagsup}=\Ccal f^{\ddagsup}
$  (Figure \ref{fig-Diracwv}a, solid and dotted curves)  
generated in $+z$- and $-z$- directions and initially also travelling in these directions at speed $\w^{j}/k^{j}=c$, with 
$$\displaylines{
\refstepcounter{equation} \label{eq-2}
\hfill 
 \vphilim^{\dagsup} =
C e^{i[k_d^{\dagsup} z-\w^{\dagsup} t +\a_0]}, \quad
  \vphilim^{\ddagsup}
=-C e^{i[-k_d^{\ddagsup} z+\w^{\ddagsup}t -\a_0] }  \hfill
 (\ref{eq-2})
}$$
(Figure \ref{fig-Diracwv} a-b, single-dot-dashed  and triple-dot-dashed curves), $\Phim=  e^{i K z}$, and $C$ ($=4C_1 /\sqrt{L}$)
a normalisation constant.   Where,
$$\displaylines{
\refstepcounter{equation} \label{eq-Doppler} 
\hfill
k^{\dagsup}=K/(1-\vel/c)=\g^{\dagsup} K,
\
k^{\ddagsup} =K/(1+\vel/c)
=\g^{\ddagsup} K
\quad {\rm and} \quad
\w^{\dagsup}=\g^{\dagsup}  \W, \
\w^{\ddagsup}=\g^{\ddagsup}\W \hfill (\ref{eq-Doppler})
}$$
  are  the source-motion  resultant Doppler-displaced wavevectors and angular frequencies; 
$\g^{\dagsup}=\frac{1}{1-\vel/c}$, 
$\g^{\ddagsup}=\frac{1}{1+\vel/c}$;
$K, \W= Kc$ are values of $k^j,\w^j$ at $\vel=0$. 
(\ref{eq-Doppler}) further gives    
 $$ \displaylines{\refstepcounter{equation} \label{eq-Kd}
\hfill k_d^{\dagsup}=k^{\dagsup}-K=\g^{\dagsup}K_d, \
k_d^{\ddagsup}=K-k^{\ddagsup}=\g^{\ddagsup}K_d \quad 
{\rm where  }\ K_d = \lf(\vel/c \rt)K. \hfill   (\ref{eq-Kd})
}$$ 
Supposing  $\vel<<c$ (yet $\vel^2/c^2$ may be large so that dynamically the  $\g$ factor in  (\ref{eq-geomean}) below can be   different from 1) 
and accordingly the de Broglie wavelength ($\lam_d =2\pi/(\g K_d)$)  later  will be much greater than the electromagnetic wavelength ($\Lam=\frac{2\pi}{K}) $,  
so 
            %the $K$ associated oscillation $\lim_{\vel<<c}\Phim =\Ci $    in $L$; that is,  
at the scale of $\lam_d$  the rapid variation of $\Ccal$ is to an external observer no different from the constant $\Ci$, that is  $\lim_{\vel<<c}\Phim =\Ci $.
Thus
$
\lf.\varphi^{\dagsup}\rt|_{\Phim =\Ci} =
 \vphilim^{\dagsup}, 
\lf. \varphi^{\ddagsup}\rt|_{\Phim =\Ci} = 
 \vphilim^{\ddagsup}
$
and the $f^{\dagsup},f^{\ddagsup}$ as given in (\ref{eq-2})
 represent external-effective space-time functions.
We see that $\vphilim^{\dagsup} $ and $\vphilim^{\ddagsup}$ are two  new plane waves travelling each to the right at equal phase velocities, $\frac{\w^{\dagsup}}{ k_d^{\dagsup} }=\frac{-\w^{\ddagsup}}{ -k_d^{\ddagsup} }=W=\frac{c^2}{\vel}$ which is  $c/\vel$ times the velocity of light $c$. 
It can be checked (\ref{Rel-E-p}) that the exact solutions $\varphi^{j}$'s and in turn the external effective $f^{j}$'s given above  
  placed in the respective wave equations  (\ref{eq-eqm1}) above and  (\ref{eq-eqmt3b}) below  
 yield exactly  the expected relativistic energy-momentum relation.

The Doppler-displaced variables 
$k_d^j$'s,$\w^j$'s in 
        %the wave functions  
the $f^{j}$'s,$\varphi^{j}$'s 
 are not single valued and thus are not good dynamical variables of the particle. The respective geometric means  
$$\displaylines{\refstepcounter{equation} \label{eq-geomean}
\hfill
k_d = \sqrt{k_d^{\dagsup}k_d^{\ddagsup}}
             %=\sqrt{(k^{\dagsup}-K)(K-k^{\ddagsup})} 
=\g K_d, 
\quad  
\w=\sqrt{\w^{\dagsup}\w^{\ddagsup}}=\g \W,  \quad 
{\rm with } \
\g=\sqrt{\g^{\dagsup}\g^{\ddagsup}}=1/\sqrt{1-\vel^2/c^2},
\hfill 
(\ref{eq-geomean})
}$$ 
 are evidently good dynamical variables of particle and  also are appropriate in view of the stochastic virtue of the electromagnetic waves.
These are also the natural independent variables of the superposed wave functions  
(for a detailed elucidation see  [\citeUnif c,d]): 
$
\widetilde{\psi}=\varphi^{\dagsup}+\varphi^{\ddagsup}
= \Ccal_d e^{i[k_d z -\w t +\alpha_0]}$ and 
$
\widetilde{\psi}{}^{{\rm vir}}
=\varphi^{{\rm vir}\dagsup}+\varphi^{{\rm vir}\ddagsup}
=-\Ccal_d^{\vir} e^{i[k_d z +\w t +\alpha_0]}
 $ which are two beat waves of a  wavelength $\lam_d=2\pi/k_d$,  travelling each at the phase speed $W=c^2/\vel$ to the right and to the left as due to the actual and reflected (virtual) charges travelling to the right and left
 respectively; 
   $\Ccal_d
=2C_1e^{i[(K+\frac{\vel}{c}k_d)z-\frac{\vel}{c}\w t]}
$,  
$\Ccal_d^{\vir}
=2C_1e^{i[(K+\frac{\vel}{c}k_d)z+\frac{\vel}{c}\w t]}
$ and  $\Ccal_d\dot{=}\Ccal_d^{\vir}\dot{=}2C_1e^{iKz}=2C_1 $ for   
$\vel<<c$.
An inspection will show that clearly   $\widetilde{\varphi},\widetilde{\varphi}{}^{{\rm vir}}$  resemble directly the de Broglie phase waves of the particle in the constant $\phi_a,\Ab$ fields here, 
 $\lam_d$ and $k_d$ are  the de Broglie wavelength and wavevector, and accordingly $\hbar k_d =p_\vel$ the linear momentum. 
 If $\phi_a,\Ab$ are arbitrarily varying in $L$ and well behaved, we can divide $L$ in a large $N$ number of small divisions in each of which the plane waves remain true and their sum gives according to the Fourier theorem the total wave, 
and  in three dimensions the de Broglie particle motion is a  straightforward extension of a locally one-dimensional motion; 
the wave equations to be given will formally  be otherwise the same (for a formal treatment in the case of a Schr\"odinger system see 
 [\citeUnif c,k]). 
We shall thus for simplicity proceed the remainder of treatment  for  constant $\phi_a,\Ab$ and, until the discussion regarding spin rotation 
in Sec. \ref{Sec-Dirac-equation},
 for a particle motion in one-dimension.

\section{Wave equation for total motion of particle
}\label{Sec-Mean}

           For single particle or for many particles without regarding the spins, the functions $\widetilde{\psi}{}$ and $\widetilde{\psi}{}^{{\rm vir}}$, or equivalently the $\fa$ and $\fb$ of (\ref{eq-tran-w1a}) below, are seen to be identical to the usual solutions to the Dirac equation, c.f. \App\ref{Dirac-solu}.  
So their wave equations, originally the (\ref{eq-eqm1}),  evidently must have a direct correspondence with the Dirac equation. The remainder of the task mainly will be to identify a physically justifiable  procedure to transform  (\ref{eq-eqm1}) 
to a form of the Dirac equation under corresponding considerations.
\begin{figure}[t]
\begin{center}
\includegraphics[width=0.8\textwidth]{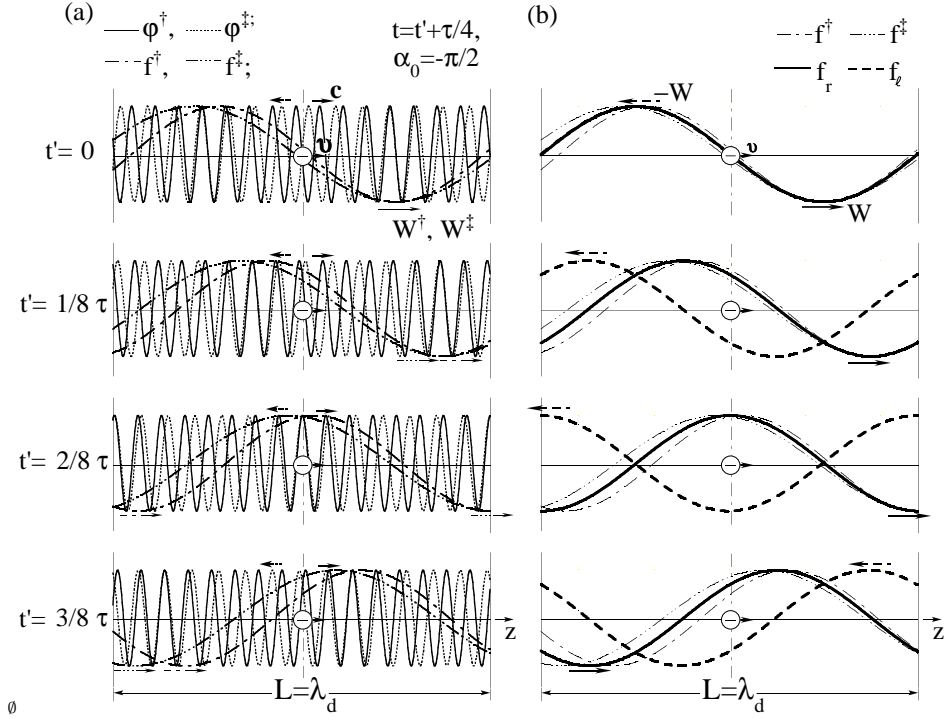}
\end{center}
\vspace{-0.6cm}
 \caption{ 
(a) shows an IED model electron
 constituted of an  oscillatory charge $-e$ of zero rest mass ($\ominus$), travelling at velocity $\vel$, and the resulting Doppler-differentiated electromagnetic waves $\vphi^{\dagsup}$ and $\vphi^{\ddagsup}$ (solid and dotted curves) of a  angular frequencies $\w^\dagsup,\w^\ddagsup$
 generated in $+z$- and $-z$- directions, 
                      % Doppler-differentiated  
plotted for the real parts in a time interval $(3/4)\tau$ in a one-dimensional box of side $L=\lam_d$; $\tau=2\pi/\w$, $\w=\sqrt{\w^\dagsup\w^\ddagsup}$, $\lam_d=2\pi((\vel/c)\w)$ being  the de Broglie wavelength.
 $\vphilim^{\dagsup}$ and $\vphilim^{\ddagsup}$ (single-dot-dashed  and triple-dot-dashed curves) are the corresponding external-effective waves, shown in both (a) and (b).  
$\fa$ and $\fb$ (solid and dashed curves) in (b) are the dynamically equivalent  mean-variable wave functions; these  resemble directly the (opposite-travelling) de Broglie phase waves and are equivalent to the space-time functions of Dirac spinor. 
}  
\label{fig-Diracwv}
\end{figure}

First, similarly as Dirac (or as alternatively  but compatibly argued in [\citeUnif c]) we want the eventual wave functions 
of the particle, and thus immediately  the $f^j$ or the original $\varphi^j$ to be linear in $\hbar\w$ and thus $\pd_t f^j$ here, so that $f^j$ at any initial time determines its value at any future time; and  we want similarly for the linear momentum here.  
We shall thus transform the second order differential  equations (\ref{eq-eqm1}) to first order ones and in the end take the limit for   $c>>\vel$
as follows.
  For the  $c^2 \nabla^2 \varphi^{j}$ terms 
of (\ref{eq-eqm1}), starting with the full wave functions $\varphi^j=\Ccal f^j$  
we first lower the spatial derivative one order as 
$\nabla^2 \varphi^j=\nabla [i \g^j K \Ccal f^j]
=i \g^j K [(\nabla \Ccal)f^j + \Ccal \nabla f^j]$. 
Restricting to $\vel << c$, we thus can 
replace $\nabla \Ccal$  by its computed value $i K \Ccal$ and
in turn put  $\Ccal\dot{=}1$ for each term; this gives
$$\displaylines{\hfill\refstepcounter{equation} 
\label{eq-deriv1}
\hfill
\nabla^2 \varphi^j|_{\Ccal\dot{=}1}
= [-\g^j K^2 \Ccal f^j + i\g^j K  \Ccal \nabla f^j]_{\Ccal\dot{=}1}
= -\g^j K^2  f^j + i\g^j K  \nabla f^j, \ \ j=\dagger, \ddagger. 
\hfill (\ref{eq-deriv1})
}$$
Next, assuming $\phi_a,\Ab$ relatively small as typically is true   in applications, 
so the resulting force constant (i.e. force per unit displacement) on the particle does not vary across $L$;   
 we can thus replace the $\nabla^2 \phiv^{j}$ in the $\phi_a,\Ab$ terms of (\ref{eq-eqm1}) by its computed value as 
$\nabla^2 \phiv^{j} |_{        \Ccal=\Ci}
=-     {{\g^j}}^2 K^2 f^{j}$, 
thus 
$$\displaylines{\hfill\refstepcounter{equation} 
\label{eq-deriv2}
\hfill\begin{array}{c}
\frac{q(\phi_a -\Ab c) }{m^{\dagsup}}\nabla^2\varphi^{\dagsup} |_{        \Ccal=\Ci} =q(-\phi_a +\Ab c)\frac{\g^{\dagsup}\W f^{\dagsup}}{\hbar}, 
\quad
\frac{-q(\phi_a +\Ab c) }{m^{\ddagsup}} \nabla^2\varphi^{\ddagsup} |_{        \Ccal=\Ci}=q(\phi_a +\Ab c)\frac{\g^{\ddagsup}\W f^{\dagsup}}{\hbar}
\end{array} \hfill
(\ref{eq-deriv2})
}$$
For the final expression we used $Kc=\W$ as earlier, and $m^j=\g^j M$ and $Mc^2 =\hbar \W $ given after  (\ref{eq-mhmu}) and (\ref{eq-geomean}). 
Finally, the $\pd_t^2 \varphi^j$'s of (\ref{eq-eqm1})
 lower one order as $\lf.\pd^2_t \phiv^{\dagsup}\rt|_{\Phim =\Ci} =-i\g^{\dagsup}\W \pd_t \vphilim^{\dagsup}$,
$
\lf. \pd^2_t \phiv^{\ddagsup}\rt|_{\Phim =\Ci} 
=i\g^{\ddagsup}\W \pd_t \vphilim^{\ddagsup}
$.
Substituting these and equations (\ref{eq-deriv1})--(\ref{eq-deriv2}) in wave equations   (\ref{eq-eqm1}), multiplying  the first resulting equation 
by $-\frac{\hbar c}{K\g^{\dagsup}} $ and  the second by $\frac{\hbar c}{K\g^{\ddagsup}}$, with 
$cK=\W$ and $\hbar K =Mc$ as before and after (\ref{eq-mhmu}), 
we eventually obtain the wave equations for the electromagnetic waves of the particle expressed by $f^{\dagsup}, f^{\ddagsup}$:  
$$\displaylines{
\refstepcounter{equation} \label{eq-eqmt3b}\label{eq-eqmt3}
\hfill
[Mc^2+q\phi_a    -  c (i \hbar \nabla +  q\Ab)] \vphilim^{\dagsup}                             =i\hbar \pd_t \vphilim^{\dagsup},                                               
 \ \  [-Mc^2+q\phi_a   + c(i \hbar  \nabla  +q \Ab )]\vphilim^{\ddagsup} =                    i\hbar   \pd_t \vphilim^{\ddagsup}.
      %%%%%%%%%%%%%%%%%%%%
      %(Mc^2+q\phi_a )      \vphilim^{\dagsup}                                         -  c (i \hbar \nabla +  q\Ab) \vphilim^{\dagsup}                             =i\hbar \pd_t \vphilim^{\dagsup},                                                \ \  (-Mc^2+q\phi_a )\vphilim^{\ddagsup}                                          + c(i \hbar  \nabla  +q \Ab )\vphilim^{\ddagsup} =                    i\hbar   \pd_t \vphilim^{\ddagsup}.
      %%%%%%%%%%%%%%%%%%%
\hfill \quad (\ref{eq-eqmt3b})
}$$

For the particle dynamics in question we want to further transform  (\ref{eq-eqmt3b}) to be expressed by the particle  wave variables, i.e. the $k_d$ and $\w$ defined in (\ref{eq-geomean}), and the corresponding wave functions, the $f_r,f_\ell$ to obtain below. We shall below obtain such  functions through a dynamic equivalence transformation directly  from the $f^{\dagsup},f^{\ddagsup}$; 
these ought to be and will show to be functions identical to the  $\widetilde{\psi}, \widetilde{\psi}{}^{\vir}$ obtained  in a physically more transparent way earlier; the present           
approach below will advantageously preserve a direct tractable connection with the original $f^{\dagsup},f^{\ddagsup}$, thus also    $\varphi^{\dagsup},\varphi^{\ddagsup}$, whose wave equations  (\ref{eq-eqmt3b}) or  (\ref{eq-eqm1}) give the  relativistic energy-momentum relation  
exactly based on the Doppler equations (\ref{eq-Doppler}), see (\App\ref{eq-releqx}) of \ref{Rel-E-p}.
 What 
(accordingly) 
is in question in the transformation mainly is to maintain an equivalence to  the quadratic equation (\App\ref{eq-releqx});
this corresponds to the equations    
              % $f^jf^{j'}=f_\mu f_{\mu'}$, 
$\frac{\pd f^j}{\pd z^\nu}\frac{\pd f^{j'}}{\pd z^\nu}
 =\frac{\pd f_\mu}{\pd {z^\nu}^n}\frac{\pd f_{\mu'}}{\pd z^\nu}$, etc., with $j,j'=\dagger,\ddagger$, 
$\mu,\mu'=r,l$, $z^\nu=t,z$  ($\nu=0,3$); and $f_\mu \frac{\pd f_{\mu'}}{\pd z}=0$.
 The equivalence condition requires in particular  the transformed quadratic to be $\frac{\pd f_r}{\pd z}\frac{\pd f_l}{\pd z}=
k_d^2$, that is, it has a plus sign in front and its cross-term product with $Mc^2$ (i.e. the $\Ocal $ discussed after \ref{eq-eqmt3bp})  is absent. 
This can be achieved if we introduce a wavevector 
being the imaginary of (thus orthogonal to) $k_d$: 
$$\displaylines{\refstepcounter{equation} \label{eq-kd2m}
\hfill
\kb_d=(\g/i \g^{\dagsup})k_d^{\dagsup}, \quad 
\kb_d=(\g/i \g^{\ddagsup})k_d^{\ddagsup}; \quad {\rm thus}\ 
k_d^{\dagsup } k_d^{\ddagsup }=(1/i^2){\kb_d}^2
\hfill(\ref{eq-kd2m})
}$$
(compare  $\kb_d$ with the operator 
%linear momentum operator 
$p_{\vel.op}=\frac{\hbar}{i}\nabla $ later). (\ref{eq-kd2m})  alternatively can be  expressed by  
$$\displaylines{
\refstepcounter{equation} \label{eq-tfdirac}
\hfill
 ({\rm a}): \  k_d^{\dagsup} (-k_d^{\ddagsup}) 
                   = 
                                \kb_d \kb_d 
\quad    {\rm or}\quad
  ({\rm b}): \ (- k_d^{\dagsup}) k_d^{\ddagsup} 
                   =
                                \kb_d \kb_d.  
\hfill (\ref{eq-tfdirac})
}
$$
We now first 
transform  the  Doppler-differentiated $f^j(;k_d^j,\w^j) $'s  (as short hand notations of $f^j(z,t;k_d^j,\w^j)$'s) to a pair of  mean (wave)-variable  functions $f_\mu(;\kb_d,\w)$'s (denoting  $f_\mu(z,t;\kb_d,\w)$'s) by, say, satisfying  (a) of (\ref{eq-tfdirac}) and ordinarily  $\w=\sqrt{\w^{\dagsup}\w^{\ddagsup}}$ of (\ref{eq-geomean}):
$$\displaylines{
\refstepcounter{equation} \label{eq-tran-w1a}
\hfill \vphilim^{\dagsup} (;k_d^{\dagsup},\w^{\dagsup})  \rightarrow  \fa(;\kb_d,\w)=C_{r} e^{i[\kb_d z -\w t+\a_0]}, 
\quad
f^{\ddagsup}  (;k_d^{\ddagsup},\w^{\ddagsup}) \rightarrow \fb(;\kb_d,\w)= C_{\ell} e^{i[\kb_d z +\w t +\a_0]}; 
                 \hfill (\ref{eq-tran-w1a})
}
$$
see these functions plotted in Figure \ref{fig-Diracwv}b.
The 
transformed $\fa,\fb$ indeed are desirably  identical functions to the original $f^{\dagsup},f^{\ddagsup}$ if disregarding  the high-order differences in the coefficients  $\g^{\dagsup}, \g^{\ddagsup} $ and $\g$ 
                   %which scale the 
in the wave variables and the reversed travel direction of $\fb$  from  $f^{\ddagsup}$. 
The $\fa,\fb$, being identical  functions  to  the 
$\widetilde{\psi}$,$\widetilde{\psi}{}^{{\rm vir}}$ earlier, indeed  are therefore the pertinent space-time  functions of the particle; 
these are each functions of the source motion and the total (electromagnetic) wave oscillation and accordingly directly resemble the de Broglie phase waves; and these are equivalent to  Dirac's  space-time functions.

To entail that in the matrix representation later (Sec. \ref{Sec-Dirac-equation}) a cross-term product $\Ocal $ discussed after (\ref{eq-eqmt3bp}) is similarly absent, for transformation of the first derivatives we are compelled to  satisfy the alternative condition (b) of  (\ref{eq-tfdirac}). The use of (\ref{eq-tfdirac}b) and ordinarily the $\w=\sqrt{\w^{\dagsup}\w^{\ddagsup}}$ of (\ref{eq-geomean}) first directly leads to the intermediate  transformations for the $f^j$'s
given in the left column  below:
$$\displaylines{
\refstepcounter{equation} \label{eq-tran-w1}
\hfill
f^\dagsup 
{\mbox{\footnotesize{}}\atop \overrightarrow{ 
\mbox{ \footnotesize{
${k_d^{\dagsup} \rightarrow -\kb_d
\atop \a_0\rightarrow -\a_0'
}$ 
}}}} 
\fb^* {\mbox{\footnotesize{}}\atop \overrightarrow{ 
\mbox{ \footnotesize{(a1) }}}} \fb, 
\quad
\nabla  \vphilim^{\dagsup}  
=i k_d^{\dagsup} \vphilim^{\dagsup}
{\mbox{\footnotesize{         }}
\atop \overrightarrow{ 
\mbox{ \footnotesize{
$ {k_d^\dagsup \rightarrow -\kb_d,    \atop
f^{\dagsup}\rightarrow \fb}$
 }}}}
i(-\kb_d) \fb
=
-\nabla \fb, \hfill 
\cr
\hfill
f^\ddagsup 
{\mbox{\footnotesize{}}\atop \overrightarrow{ 
\mbox{ \footnotesize{
${k_d^{\ddagsup} \rightarrow \kb_d
\atop 
\a_0\rightarrow \a_0'}$
 }}}} 
               %-C_1 e^{i[-\kb_d z+\w t +\a_0']}=
-\fa^* {\mbox{\footnotesize{}}\atop \overrightarrow{ 
\mbox{ \footnotesize{(a2) }}}} -\fa, 
\quad
\nabla  \vphilim^{\ddagsup}
=-ik_d^{\ddagsup}\vphilim^{\ddagsup}
                  {\mbox{\footnotesize{}}\atop \overrightarrow{ 
\mbox{\footnotesize{
${k_d^\ddagsup\rightarrow \kb_d, \atop
f^{\ddagsup}\rightarrow -\fa
}$
}}}}
-i\kb_d (-\fa) 
=\nabla  \fa \hfill        
   (\ref{eq-tran-w1})
}$$
where, $\fb^*=C_{\ell} e^{i[-\kb_d z-\w t -\a_0']}$, 
$\fa^*=C_{r} e^{i[-\kb_d z+\w t +\a_0']}$, $\a_0'=-\a_0$;
the transformations (a1) and (a2) in (\ref{eq-tran-w1})  finally naturally lead to the same $\fb, \fa$ as in (\ref{eq-tran-w1a}),  which indeed  
also  represent the original $\fb^*,\fa^*$ in all aspects (having 
the same 
phase velocities 
          %($-\w/\kb_d=-(-\w)/(-\kb_d)=-c^2/\vel$ and $\w/\kb_d=-\w/(-\kb_d)=c^2/\vel$ and)  
           %
and wave forms) except the opposite rotating phases  on the complex plane 
and the opposite signs of $\a_0'$ and $\a_0$ that  altogether are dynamically inconsequential. The relations in the left column and the use of  (\ref{eq-tfdirac}b) again then lead to the results in the right column which is actually in question in respect to dynamical equivalence here.

Substituting in wave equations (\ref{eq-eqmt3b}) the transformation relations (\ref{eq-tran-w1a}) and   (\ref{eq-tran-w1}) gives
$$\displaylines{
\refstepcounter{equation} \label{eq-eqmt3bp}
\label{eq-eqmt3p}
\hfill
(Mc^2+q\phi_a)    \fa
   +c (i \hbar  
\nabla   -q\Ab)\fb
=i\hbar \pd_t \fa,
\ \
 (-Mc^2+q\phi_a)\fb
   + c(i \hbar  
\nabla  +q\Ab)\fa
=
i\hbar   \pd_t \fb. \ \ 
\hfill (\ref{eq-eqmt3bp})
}$$
As a check, placing in (\ref{eq-eqmt3bp}) the $\fa, \fb$ of  (\ref{eq-tran-w1a}), multiplying the first and the negative of the second resulting equations, dividing $\fa\fb$, putting for simplicity $\phi_a=A_a=0$ for the problem mainly of concern here,  we correctly obtain the same result as (\App\ref{eq-releqx}):
%the relativistic energy-momentum relation, 
$ M^2c^4  -\hbar^2  \kb_d^2 c^2  +\Ocal =\hbar^2 \w^2 $ where  $\kb_d^2 =-k_d^2$ following (\ref{eq-kd2m}) and (\ref{eq-geomean});   $\Ocal=0$. 
%This gives .

\section{Two-particle system: spins, charges, and time-arrows}\label{Sec-Spin}

Consider two spin-half  particles 1,2 
having identical sets of $\{\fa, \fb\}$'s 
tend to occupy the same location $z$ or more precisely  the  same region in ($0,L$); suppose these are  noninteracting (a finite particle-particle interaction can in principle be included in $V^j$ and will not affect the general conclusions below).
             %*******
\begin{figure}[b] %[htpb]
%\vspace{0.2cm}
\begin{center}
\includegraphics[width=0.80\textwidth]{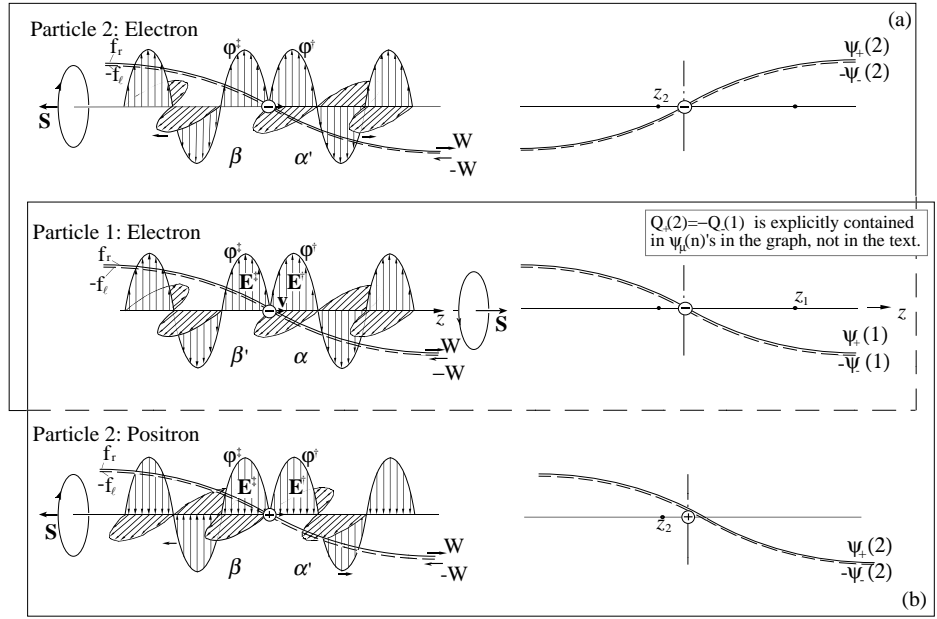}
\end{center}
\vspace{-0.8cm}
 \caption{ 
Two spin-half IED particles 1 and 2 described by identical, thus mutually symmetric sets of  Doppler-displaced space-time functions $\{\varphi^{\dagsup},\varphi^{\ddagsup}\}$'s or effectively 
$\{\fa, \fb\}$'s   tend to occupy the same location $z$;
 panel (a) shows  two electrons and  (b)  an electron and positron. 
Particle 1 has a spin  $\Sb$ in $+z$- direction,  denoted spin up, $\ai$, and is parallel with  the generating direction of the $\fa$ wave; particle 2 has  spin $\Sb$ in $-z$-direction, denoted spin down $\bii$, and is  parallel with $\fb$;
their opposite counterparts (termed virtual spins) $\bi$ and  $\aii$ are parallel with  $\fb$ and $\fa$. 
Right graphs show the complete wave functions $\{\psipi, \psimi \}$ and $\{\psipii, \psimii \}$ of particles 1 and 2. In panel (a),
$\psi_{\nu}(1)$ and $\psi_{\nu}(2)$ ($\nu=+,$-) are antisymmetric  due to the antisymmetric spin functions and same charges. In (b),
these are symmetric (as shown in the graph) due to antisymmetric spins and also  opposite charges ($\Qcal_\p(2)=-\Qcal_\m(1)$),  the latter of which leads to the radiation electric fields ${\bf E}^j$'s are opposite in direction  (in the text the $\Qcal_\p(2), \Qcal_\m(1)$ are not explicitly regarded, rendering the $\psi_{\nu}(1)$ and $\psi_{\nu}(2)$ for the  electron and positron  to be antisymmetric and the same as for two electrons).
}  
\label{fig-eldirac}
\end{figure}
In virtue of the statistical nature of the electromagnetic displacements, the probability of finding a portion of  particles 1 and 2 at  locations $z_1 $ and $z_2$ is proportional to the product $f_\mu(z_1,t)f_{\mu'}(z_2,t)$. 
Since  these have identical space-time function sets,
the corresponding total space-time function  is evidently symmetric, thus $f_{s} $, here in the only form  
$f_{s}(z_1,z_2) =\frac{1}{\sqrt{2}}[\fa(z_1,t)\fb(z_2,t)+\fb(z_2,t)\fa(z_1,t)]$ to be compatible with the antisymmetric total spin function later. 

In the case of  two identical electrons (Figure \ref{fig-eldirac}, panel a),  their spins then need  according to Pauli principle be opposite (left graph in the figure) to avoid both particles occupying the same quantum state; the  total spin function  for this is antisymmetric ($\chi_a=\frac{1}{\sqrt{2}}[\ai\bii-\aii\bi]$).
Apparently it is in general also relevant that we introduce charge functions $\Qcal_{\m} (1),\Qcal_{\m}(2)$  to reflect the sings of charges 1,2; these for the two like charges are  trivial identities and lead to a trivial  symmetric total charge function, $\Qcal_s$. 
     The above functions together define an antisymmetric   two-electron function (Figure \ref{fig-eldirac}a, right graph),
$\psi_a
                 %(z_1, \ai; z_2, \bii)
=\vphilim_s \Qcal_s\chi_a$, yielding as expected a total probability independent of how we sample the two stationary-state identical,   indistinguishable (as result of being identically extensively distributed in $L$) particles.

In the case of an electron and  positron (Figure \ref{fig-eldirac}, panel b), the total two-particle function is symmetric (right graph), thus $\psi_s$, as follows from the   observational fact  
that two such particles can approach each other arbitrarily close and, in an extreme case  annihilate into "one point" in the vacuum. 
The total charge function for their two opposite signed charges evidently needs   be antisymmetric, thus $\Qcal_a$. 
(If imagine the wave displacement $\varphi^j$  in the medium is executed by a chain of dipole charges then it is immediately clear that the corresponding radiation electric field $E^j$,  with  $|E^j| \propto \varphi^j$, produced by the positive charge is reversed from that by the negative charge, see  left graph in 
Figure \ref{fig-eldirac}b.)  
Placing the two known functions in  
$\psi_s=\Qcal_a \vphilim_s \chi_{_{{\rm Sym}}}$ 
gives therefore  Sym=antisymmetric and thus $\chi_{_{{\rm Sym}}}=\chi_a$
for the electron-positron system.

Our particles contain each the electromagnetic waves $\varphi^{\dagsup},\varphi^{\ddagsup}$, or effectively 
$\fa,\fb$,    
 generated   in the specified $+z$- and $-z$- directions which 
 assume definite relationships 
with the  spin orientations that turn out in a measurement: 
          Given, say, particle 1 is measured to be  spin up, $\ai$, with   $\fa$ being parallel with it, then its other internal process, $\fb$, is  parallel with a virtual (indicated by a prime) spin-down state, $\bi$. Similarly for particle 2, being then actually spin down, thus  $\fb$ parallel with $\bii$, its $\fa$ is parallel with a virtual spin-up state, $\aii$.
 From the foregoing antisymmetric spin requirement  follow the relations for the spin functions:  
$$\displaylines{
\refstepcounter{equation} 
\label{eq-spinx2}
\hfill
\aii=-\ai, \quad \bi=-\bii;
\quad {\rm with } \ \ 
 \bi =-\ai, \quad \aii=-\bii 
\hfill (\ref{eq-spinx2})
}$$
following the opposite signs as is meant by  the "virtual" spins.
Through  (\ref{eq-spinx2}), 
the virtual spin vector of particle 1, $\Sb'(1)$ (virtual spin down) and the actual spin vector of particle 2, $\Sb(2)$ (actual spin down), pointed each in the $-z$- direction, are now each represented as scalar quantities with minus signs. 
%  each relative to the   $+z$-direction. 
               %%%below go************

If disregarding the signs of charges explicitly, 
the electron-electron and the electron-positron are    
two equivalent systems of identical, spin-half particles,  
each described by the space-time functions $f_\mu(;\kb_d,\w)$'s  
with $\mu=r,\ell$,        
which joined together with the spin functions give the complete wave functions: $\psi_{\nu}(n;\kb_d,\w)=f_{\mu}(;\kb_d,\w) \a_{\nu}(n)$ with $\nu=+$,-, $n=1,2$, 
 $\a_\nu=\a,\beta$. 
Now as a further step to conform our wave equation later to matrix form, 
we hereafter require that the $\psi_{\nu}(n)$ functions are  elements of a matrix of one column, $\psib$.   
The matrix wave equation itself will entail the two desired features discussed after equation (\ref{eq-eqmt3bp}), that is,  (i) the product  of $-k_d$ and $k_d$ in the quadratic equation is positive: $ k_d^2$ (entailed by the situation that these in the matrix form are offdiagonal elements, see  (\ref{eq-eqmt3pp-el2}) or (\ref{eq-matrx1}), and (ii) the total cross-term product  $\Ocal=0$ (entailed by the characteristics that in matrix equation the $\psi_{\nu}(n)$'s are explicitly mutually orthogonal).
The first of these two features which we have up to now enforced by use of $\kb_d$ for $k_d$, should no longer be used  in the matrix form to avoid a dual accounting. The space-time functions accordingly write  
 $ \fa(;k_d,\w)=C_{r} e^{i[k_d z-\w t]}$, 
$\fb(;k_d,\w)=C_{\ell} e^{i[k_d z+\w t]}$
with  $k_d$ the ordinary scalar quantity and related with $k_d^{\dagsup},k_d^{\ddagsup}$  through (\ref{eq-geomean}).  
Accordingly,
 $$\displaylines{\refstepcounter{equation} \label{eq-spins2}
\psipi=\ai\fa (;k_d,\w)
=\ai C_{r}  e^{i[k_d z- \w t]},
\ \
\psimii=\bii\fa (;k_d,\w)
=\bii C_{r}  e^{i[k_d z- \w t]},
\hfill (\ref{eq-spins2}a)
\cr
   \psimi
= \bi\fb (;k_d,\w)
=\bi C_{\ell}  e^{i[k_d z+ \w t]};
 \  \
\psipii=\aii\fb (;k_d,\w) 
=\aii C_{\ell}  e^{i[k_d z+ \w t]}.
\hfill (\ref{eq-spins2}b)
}$$
The complete wave functions (\ref{eq-spins2}a)--(b) 
(Figure \ref{fig-eldirac}, right graphs) describe  
two identical particles of opposite oriented actual spins and accordingly opposite virtual spins, and  these are identical to  
the solutions (see equation \ref{eq-Dx1}) for  Dirac equation.
We can readily check that placing  the foregoing relations in the antisymmetric total function for two spin-half (like charge) particles, $\psi_{a} =\frac{1}{\sqrt{2}}[\psipi\psimii-\psimi\psipii]$  correctly leads to that the probability of finding two identical particles at any location $z$ in $L$ is not altered by 
interchanging the locations of the particles (the indistinguishability). 
We can also check that the same  $\psi_{a}$ is given by the product of the separate total functions: $\psi_{a}(z_1,z_2)=f_{s}(z_1,z_2)\chi_a(1,2)$; notice that once we specified say particle 1 is spin up and 2 spin down, 
then $\fa(z_1,t)\fb(z_2,t)\bi\aii$ and $\fb(z_1,t)\fa(z_2,t)\ai\bii$
are zero 
since these do not describe the present reality. 

Lastly, the spin-up state of particle 1, $\ai$, is associated with 
an effective electromagnetic wave  
 $\fa$ travelling to the right, thus  $ \pd_t \fa/\fa =-i \w $, 
while  the spin-up state of particle 2  with $\fb$ travelling to the left, thus $\pd_t \fb /\fb= i \w  $; the latter has as if a reversed time arrow relative to the former. 
We may  introduce the time arrow functions defined for particles 1 and 2 as
$
            %$\displaylines{\refstepcounter{equation} \label{eq-space-spin3}\hfill 
 \Tcal(1)=1, 
            %\quad 
\Tcal(2)=-1,  
          %\hfill (\ref{eq-space-spin3})}$
$
such that the action of these on the time derivatives project the wave propagations to be  both 
in the 
$+z$-direction: $\Tcal(1) \pd_t \psi_{\nu}(1)=\pd_t \psi_{\nu}(1)$,
$
 \Tcal(2)\pd_t \psi_{\nu}(2)= -\pd_t \psi_{\nu}(2)$.

\section{Dirac equation}\label{Sec-Dirac-equation}

For two identical, spin-half particles of identical sets of space-time functions $\fa, \fb$ described by wave equations (\ref{eq-eqmt3bp})
 tending to occupy the same  location $z$, 
 we shall now express the corresponding wave equations in terms of the complete wave functions of Sec. \ref{Sec-Spin}.
For particle 1, we thus  multiply the first equation of (\ref{eq-eqmt3bp}) by $\ai$  and the second    
by  $\bi$,  
act   $\Tcal(1)$  in front of the time derivatives, denote its charge by  $q_1$, 
and  get  
$$\displaylines{\refstepcounter{equation} \label{eq-Diraceqa}
\hfill
(Mc^2 +q_1\phi_a )   \fa\ai
   + c(i \hbar  
\nabla -q_1\Ab)\fb (-\bi ) 
=i\hbar \Tcal(1) \pd_t \fa \ai,   \qquad
\hfill
\cr
\hfill 
 (-Mc^2+q_1\phi_a)\fb\bi 
   + c (i \hbar  
\nabla +q_1\Ab)\fa    (-\ai) 
=
i\hbar  \Tcal(1) \pd_t (\fb \bi).
\hfill (\ref{eq-Diraceqa})
}$$
For particle 2, instead we  multiply  the first equation of (\ref{eq-eqmt3bp})  by $-\bii $ and the second  by $-\aii$, act both  equations by $\Tcal (2)$, denote its charge by $q_2$, and  get 
$$\displaylines{\refstepcounter{equation} \label{eq-Diraceqb}
\hfill
  (Mc^2+q_2\phi_a)\fa (-\bii)    + c (i \hbar  \nabla -q_2\Ab)\fb (+\aii ) =i\hbar \Tcal(2)\pd_t \fa (-\bii), \qquad
\hfill
\cr
\hfill 
% (Mc^2-q_2\phi_a)\fb\aii    -  c (i \hbar \nabla +q_2\Ab) \fa     (-\bii)  = -i\hbar\Tcal(2)   \pd_t \fb \aii. \cr
 (-Mc^2+q_2\phi_a)\fb(-\aii)    +  c (i \hbar \nabla +q_2\Ab) \fa     (+\bii)  = i\hbar\Tcal(2)   \pd_t \fb (-\aii). 
\hfill (\ref{eq-Diraceqb})
}$$
 In the second terms  in equations (\ref{eq-Diraceqa})--(\ref{eq-Diraceqb})  we made the replacements  
$\ai\rightarrow -\bi $ and 
$\bi \rightarrow -\ai$, $\bii \rightarrow -\aii$, 
$\aii \rightarrow -\bii$ based on  
 (\ref{eq-spinx2}), 
to conform to the transformed space-time functions earlier.

Substituting in (\ref{eq-Diraceqa})--(\ref{eq-Diraceqb}) with  (\ref{eq-spins2}a)--(b)  for the $\psi_{\nu}(n)$'s and  the 
   %(\ref{eq-space-spin3}) for 
$\Tcal(n)$'s expressed earlier,  
and, to form a  direct contrast between the actual spin directions of the two particles, 
re-arranging the resulting four  equations in the order of   spin-up states of particles 1 and 2 first and then 
 spin-down states of particles 1 and 2, 
we finally obtain a set of four coupled linear  first order  partial differential equations governing the motions of the two  particles in terms of $\psi_{\nu}(n)$:
$$\displaylines{
\refstepcounter{equation} \label{eq-eqmt3pp}
\label{eq-eqmt3pp-el2}
\hfill\quad  
(Mc^2+q_1\phi_a)      \psipi
   - c(i \hbar  
\nabla -q_1\Ab)\psimi      
=i\hbar \pd_t \psipi  \quad (\mbox{particle 1, spin up})
\hfill \quad\qquad
\cr
       %********
\hfill\quad 
(Mc^2-q_2\phi_a)      \psipii
   +  c (i \hbar  
\nabla +q_2\Ab)
  \psimii    
=i\hbar \pd_t \psipii
\quad (\mbox{particle 2, spin up})\hfill \quad\qquad
\cr
%******
\hfill
  (-Mc^2+q_1\phi_a) \psimi
   -  c (i \hbar  
\nabla +q_1\Ab)
  \psipi
=
i\hbar   \pd_t \psimi 
\quad (\mbox{particle 1, spin down})\hfill \qquad
\cr
         %*******
\hfill
  (-Mc^2-q_2\phi_a)\psimii
   +  c (i \hbar  
\nabla -q_2\Ab)
  \psipii 
=
i\hbar   \pd_t \psimii \quad (\mbox{particle 2, spin down})
\hfill 
 (\ref{eq-eqmt3pp-el2})
}$$
From the discussion of Sec. \ref{Sec-Spin} that the $\psi_{\nu}(n)$'s and accordingly also their first derivatives 
 are mutually orthogonal, it follows that the linear 
equations  (\ref{eq-eqmt3pp-el2}) are equivalent to a matrix equation. Supposing specifically the two particles are a positron and an electron and therefore    $q_1=q$, $q_2=-q$, 
the matrix form of  (\ref{eq-eqmt3pp-el2}) is thus   
$$\displaylines{
\refstepcounter{equation}
\label{eq-eqmt3ppp}\label{eq-DiracMeq}
\hfill
  H_{op}\psib= i\hbar \pd_t \psib, \ \
{\rm with  } \ H_{op}=\betamt  Mc^2+ q\phi_a   + c \alphab (\pbf_{\vel.op }-q\Ab) \ {\rm and} \ \pbf_{\vel.op}=-i\hbar  \nabla              \hfill             (\ref{eq-DiracMeq})
}$$
being the relativistic total Hamiltonian
and linear momentum operators. Where, 
$$\displaylines{
\hfill
  \betamt=\left(
\begin{array}{cc}
{I} & 0  \cr
0 & -{I} 
\end{array}
\right),\quad
\alphab=\left(
\begin{array}{cc}
0 & {\pmb{\sigma}}_z  \cr
{\pmb{\sigma}}_z  & 0 
\end{array}
\right), 
\quad
{\pmb{\psi}}=
\left(
\begin{array}{c}
{\pmb{\psi}}_{+}  \cr
{\pmb{\psi}}_{\m}   
\end{array}
\right) \quad {\rm and}  \hfill
\cr
\hfill
{\Imtr}=\left(
\begin{array}{cc}
1 & 0  \cr
0 & 1
\end{array}
\right); 
            \quad
            {\pmb{\sigma}}_z=
            \left(
            \begin{array}{cc}
            1 & 0  \cr
            0 & -1
            \end{array}
            \right)
\quad
{\pmb{\psi}}_{+}=
\left(
\begin{array}{c}
\psipi  \cr
\psipii   
\end{array}
\right); \quad    
{\pmb{\psi}}_{\m}=
\left(
\begin{array}{c}
\psimi\cr
\psimii
\end{array}\right). \quad 
  \hfill 
\refstepcounter{equation} \label{eq-paulimatcomp} 
(\ref{eq-paulimatcomp})
}
$$
The off-diagonal  elements of the matrix $\sigmab_z$, $\sigma_{z11}(=1)$ and $\sigma_{z22}(=-1)$ here correspond to the $\ai=1$ and $\aii=-1$ earlier. 
We see that, 
${\pmb{\psi}}$ is equivalent to a Dirac spinor,    
 ${\sigma}_z$  
the $z$-component of  Pauli matrices, and as a whole, equation  (\ref{eq-eqmt3ppp}) 
              %or   (\ref{eq-eqmt3pp-el2})  
is identical to the  Dirac equation for an electron-positron system equivalent to here.  
For the present  case rotation transformation is trivial, so  $\vec{\sigma}=
%0 \hat{x}+0 \hat{y} +
\sigma_z  \hat{x}$. 
             %($ S_{z}= \frac{\hbar}{2} {\sigma}_z $ and ${\mu}_{sz}=- e S_{z} /m$ give the spin angular momentum and magnetic moment.) 

Suppose  more generally the two particles' spin angular momenta, $\Sb$ ($=\frac{\hbar}{2} {\sigmab}$)'s, are along an axis ${\bf{n}}$ 
 executing in general a   precession about the $z$-axis at a fixed angle ($\arccos (\frac{S_z}{S})$).  
For each particle  being in stationary state,  its $\Sb$ (similarly its magnetic moment $\pmb{\mu}_s   (=- e \Sb /m)$)  as a vector quantity when in small rotations about 
the $z$-axis must maintain invariant with respect to 
its projection on the $z$-axis, ${\bf n} \cdot \Sb =\pm \frac{1}{2}\hbar$, 
and is Hermitian. 
In addition to the antisymmetric condition given by the 
$\sigma_z$ of  (\ref{eq-paulimatcomp}) above,  
an infinitesimal rotation transformation as such needs be  unitary.  
A specific  set of transformation matrices having these  properties are known to be the Pauli matrices, $\sigma_x, \sigma_y$ and $\sigma_z$ of the standard expressions and 
$\sigma_z$ as expressed in  (\ref{eq-paulimatcomp}), 
$\sigmab=\sigmab_x \hat{x}+\sigmab_y \hat{y}+\sigmab_z \hat{z}$.
%%%
% and $\sigma_x, \sigma_y$ of the standard expressions. 
%and  
%              its $x$ and $y$ components being: 
%         $$\displaylines{
%         \refstepcounter{equation} \label{eq-Paulim}
%         \qquad
%         {\sigma}_x=\left(
%         \begin{array}{cc}
%         0 & 1  \cr
%         1 & 0
%         \end{array}
%         \right),
%                     %*******
%         \quad
%         {\sigma}_y=\left(
%         \begin{array}{cc}
%         0 & -i  \cr
%         i & 0
%         \end{array}
%         \right).
%         \hfill (\ref{eq-Paulim})
%         }$$
And, the unitary matrix $I \hat{z}$ about the $z$-axis  naturally extends to a   unitary matrix  about the new ${\bf n}$- axis in three dimensions, given by  $\vec{\Imtr}=\Imtr\hat{x}+\Imtr\hat{y}+\Imtr\hat{z}$. 
Substituting in (\ref{eq-eqmt3ppp}) with $\sigmab$ and $\Imtr$ for   $\sigmab_z$ and $I$ gives a Dirac equation of the same form, now for spins in arbitrary directions. 
          %It is straightforward to extend the Dirac equation to for particle motion in three dimension following similar discussion as in [\citeUnif c].

\setcounter{equation}{0}
\setcounter{section}{0}

\section*{Appendixes IA-IC:}
\begin{appendix}

 \section{Total energy and inertia of particle wave
}\label{Sec-wave-energy}

As a general result of classical electrodynamics  based on   solution to the Maxwell's equations combined with  Lorentz force law, an electromagnetic wave $j$ transmits at the speed of light $c$ a  wave energy $\eng^{j}$ and  a linear momentum  
$
p^j=\eng^j/c$. 
Here, the amplitudes of  $\eng^j$, accordingly of $p^j$, $E^j, B^j$ and $\varphi^j$, etc., 
are  continuous values. 
Following 
M. Planck's discovery of quantum theory  in 1901, it has been 
additionally understood  
that these quantities are    {\it by nature}  quantized in amplitudes;
an electromagnetic wave of frequency $\w/2\pi$ has an energy     $\eng=n \hbar \w$, 
consisting in general of $n$ momentum-space quanta, 
or photons,
each of an energy $\hbar \w$;
and the classical continuous amplitude solutions to these are only approximations when $n$ is large.  
In the present problem, in conformity with experiments,  especially  the   pair processes,  the electromagnetic wave comprising our basic particle has  a "single energy quantum", $n=1$; 
so $\eng=\hbar \w$.
It has been further  proven especially through quantum electrodynamics that the Maxwell's equations,  and the subsequent classical wave equation (\ref{eq-CMwave1}) or  (\ref{eq-eqm1}), 
continue to hold, and the quantisation of the fields and  wave energy etc. is the result of subjecting the canonical displacement and momentum, the $u(=a\varphi)$ and  $\dot{u}$ here, to the quantum commutation relation $[u,\dot{u}]=i\hbar$.

 The total  wave of our particle of a single "quantum energy level" $\hbar \w$ in a one-dimensional box has, following  the solution to the Maxwell's equations earlier [see after (\ref{eq-CMwave1})],  two components, $\varphi^{\dagsup}$ and $\varphi^{\ddagsup}$, 
with their frequencies being 
 Doppler-displaced to  $\w^{\dagsup} $ and $\w^{\ddagsup} $  as a result of the source motion as given in (\ref{eq-Doppler}), which are related to $\w$ through (\ref{eq-geomean}).  
For the total wave comprising the particle, $\eng$   represents therefore a dynamical variable of the  particle, here the total energy of the  particle. 

The 
electromagnetic waves,   ${E^j,B^j}$'s or $\varphi^j$'s, 
 rapidly oscillating at frequencies $(\w^j/2\pi)$'s, of a geometric mean frequency $\w/2\pi$ and wavelength $\lam=c/(\w/2\pi)$ will, when ignoring the detailed oscillation as will effectively manifest at some distance, appear as if being two  rigid objects, wavetrains,  travelling at the speed of light $c$.
In view that their speed  of travel, $c$, is {\it finite} as contrasted to infinite, the wavetrains have inevitably each  {\it finite}  
inertial masses, $m^j$'s, thus an inertial mass  $m=\sqrt{m^{\dagsup}m^{\ddagsup}}$ for the total wavetrain            
and hence its resulting particle.                
This mechanical depiction of the total wave, as a rigid "wavetrain", permits us at once to  express 
according to  Newtonian mechanics 
the linear momentum of the wavetrain to be $p=mc$. 
Combining this with the classical electrodynamic result  $\eng =pc$ above 
gives  the kinetic energy of the wavetrain $\eng =mc^2  $, being equivalent to the Einstein's mass-energy relation.
  This energy and the Planck energy earlier ought  to equal, thus 
$$\displaylines{\refstepcounter{equation}\label{eq-mhmu}\hfill 
m= \hbar \w /c^2; \quad      
                   %{\rm or \putting } \ m= \g M, \quad
\mbox{ or at $\vel=0$:} \ M=\hbar \W /c^2
\hfill (\ref{eq-mhmu})}$$
with $M$ the rest mass of the particle; combining (\ref{eq-mhmu}) with  (\ref{eq-geomean}) gives 
$m=\g M$. Combining   (\ref{eq-mhmu}) with $p=mc$ further gives $mc=(\hbar \w/c^2)c=\hbar k$ and accordingly $M c =\hbar K$, with $\w =k c$, $\W =K c$ and $k=\g K$ as earlier.

\section{Relativistic energy--momentum relation for the electromagnetic waves of particle
} \label{app-A}\label{Rel-E-p}

Consider first the simpler case of $A_a=0$.  Placing in wave equations (\ref{eq-eqmt3b}) with  
$\vphilim^{\dagsup},\vphilim^{\ddagsup } $ of  (\ref{eq-2}),
dividing the resulting first and second equations  by  $\vphilim^{\dagsup}$ and $-\vphilim^{\ddagsup}$ and 
sorting give
$
 Mc^2    + \hbar k_d^{\dagsup}
c
                                  =\hbar \w^{\dagsup} -q\phi_a$, 
$ 
Mc^2   -  \hbar k_d^{\ddagsup}
c
                                 =\hbar \w^{\ddagsup}+q\phi_a $. 
Multiplying  gives  
$$\displaylines{
\refstepcounter{equation} \label{eq-relmen}
\hfill
M^2c^4 - \hbar^2 k_d^{\dagsup}{}k_d^{\ddagsup}{}c^2 +\Cross=\hbar^2 \w^{\dagsup}\w^{\ddagsup}-q^2\phi_a^2    
              \hfill(\App\ref{eq-relmen})
}$$
Where 
$k_d^{\dagsup} k_d^{\ddagsup}
=k_d^2$ and $\w^{\dagsup}\w^{\ddagsup} =\w^2$ following  (\ref{eq-geomean}); 
$Q=Mc^2 c\hbar (k^{\dagsup}_d{}-k^{\ddagsup}_d{}) 
$,  with 
$k_d^\dagsup{}-k_d^\ddagsup{}=2k_d (\frac{\vel}{c})\g $
and $ Mc^2=\hbar K c$, so $Q=2\hbar^2 k_d{}^2 c^2$. 
With these, putting $\hbar k_d=\pm p_\vel$, 
$\hbar \w=\pm \eng$ where  
$p_\vel$,$\eng$ are here variables having positive and negative solutions and thus the right hand side of    (\App\ref{eq-relmen}) reduces as 
$\sqrt{[ (\hbar \w -q\phi_a)(\hbar \w+q\phi_a)]^2}
=\sqrt{[- (\eng -q\phi_a) (\eng-q \phi_a)]^2} 
= (\eng -q\phi_a)^2$, then  
(\App\ref{eq-relmen}) reduces exactly to 
$M^2c^4 +p_\vel^2 c^2=(\eng -q\phi_a)^2$.   This, or this in  the more familiar form for $\phi_a=0$, 
$$\displaylines{  \refstepcounter{equation} \label{eq-releqx}\label{eq-releq3}\label{eq-releq2}
\hfill
M^2c^4 +c^2p_\vel^2=\eng^2,
 \hfill (\App\ref{eq-releqx})
}$$
  gives just the experimentally widely corroborated relativistic energy-momentum relation.  
For the more general case of $\Ab$ finite, 
 denoting  
$k_d^{\dagsup}{}'
=\kb_d^{\dagsup}-\frac{q\Ab}{\hbar}
$, $k_d^{\ddagsup}{}'=k_d^{\ddagsup} +\frac{q\Ab}{\hbar}
$,  
the particular feature  that (the effective portion of) $\Ab$ is always perpendicular to $k_d \zb$ leads to 
$k_d{'}^2
=k_d^{\dagsup}{}' k_d^{\ddagsup}{}'
=k_d^2 -q^2\Ab^2/\hbar^2$,
or, $(\pm p_\vel'{})^2\equiv (\pm \hbar k_d')^2
=\mp (\hbar k_d -q \Ab )(-\hbar k_d -q \Ab)
=[\mp (\pbf_\vel -q \Ab )]^2$.
(\App\ref{eq-releqx}) thus generalises to $M^2c^4 +c^2p_\vel{'}^2=(\eng-q\phi_a)^2$.

\section{Solution of Dirac equation from the standpoint of particle internal process  
}
\label{App-Fvec}
\label{Dirac-solu}

We shall here  mainly discuss  the choice of the solution forms of  the Dirac equation from the standpoint of internal processes of the IED particle model for simplicity for  spins along $z$-axis, in an otherwise  basically standard procedure. The two equations of (\ref{eq-Diraceqa}) or (\ref{eq-Diraceqb}) for particle $n=1$ or 2 are coupled in $\psi_{\p}(n)$ and $\psi_{\m}(n)$ and can not be solved separately as in   \ref{Rel-E-p}.  We need to solve each two, or more generally the four equations of the Dirac equation (\ref{eq-Diraceqa}) together. Let the trial functions be:
$\psi_{\nu}(n)=C_{sn} e^{\frac{i}{\hbar}[p_\vel  z- \eng t ]}$, $ s=+,-$, $n=1,2$.
Placing these in  (\ref{eq-DiracMeq}) and rearranging give 
$$\displaylines{\refstepcounter{equation} \label{eq-matrx1} 
\left(
\begin{array}{cccc}
\eng-Mc^2-q\phi_a &0&-[\pbf_\vel-q\Ab] c&0  \cr
0&\eng-Mc^2-q\phi_a&0&[\pbf_\vel-q\Ab] c \cr
-[\pbf_\vel-q\Ab] c&0&\eng+Mc^2-q\phi_a&0  \cr
0&[\pbf_\vel-q\Ab] c&0&\eng+Mc^2-q\phi_a  
\end{array}
                       \right)
 \left(
\begin{array}{c}
\psipi  \cr
\psipii  \cr
\psimi  \cr
\psimii  
\end{array}
\right)      =0
\hfill
\cr 
\quad\hfill (\ref{eq-matrx1})
           %**************
}$$
(\ref{eq-matrx1}) corresponds to the four linear, homogeneous algebraic equations (\ref{eq-flineq}) for the $\psi_{s,j}$'s as four unknowns  below;  for these to have nontrivial solutions, the determinant for the matrix of the coefficients of (\ref{eq-matrx1}) needs be zero. This   
 is  
det$=[(\eng-q\phi_a)^2-M^2c^4]^2-c^4 \pbf{'}^4=0 $, with 
$\pbf_\vel'=\pbf_\vel-q\Ab$. 
                        %here $q=\pm q_1, \mp q_2$. 
This has two degenerate sets of 
 square roots solutions:   
 $
\eng-q\phi_a=\pm \sqrt{M^2c^4 + p_\vel{'}^2 c^2}
$,
which being  identical to (\App\ref{eq-releq3}). 
In view that each particle has internal processes, we thus naturally assign symmetrically two of the four solutions to particle 1, 
as $ \eng-q\phi_a =\pm\sqrt{M^2c^4+p_\vel{'}^2 c^2 }$,
 and the other two for particle 2 as 
$
\eng-q\phi_a =\mp\sqrt{M^2c^4+p_\vel{'}^2 c^2 }
$.
These two distinct sets of square-roots solutions
to the algebraic equation above represent two (distinct, identical) particles, like an electron and a positron, which do not transit from one to the other, a point agreeing  with reality and having been stressed by P.A.M. Dirac from the very beginning in  \cite{Dirac1928A}. These algebraic solutions are in contrast to the usual problem of eigen values arising generally from boundary conditions and being each  possible states of same particle between which transitions generally can occur.  
 
%%%%%%%%%%%%keep below:******************
% The Dirac particles, and hence equation (\ref{eq-App-EP0}) can on the  other hand similarly be subject to small boundary conditions and thus the energy and momentum solutions must be in general quantized, which gives the usual quantum mechanical eigen value problem for the Dirac equation.  We also bear in mind that, where Dirac equation is applicable is often high (kinetic) energy  and involving no small boundaries, in such cases one expects quantisation  is not a pronounced problem.  
%*********************************************

With $p_\vel$ and $\eng$ as known parameters, we  further solve the four algebraic equations
$$\displaylines{\refstepcounter{equation} \label{eq-flineq}
\hfill (\eng-(Mc^2+q\phi_a)) \psipi =p_\vel'c \psimi  \quad (a), 
\quad
(\eng-(Mc^2+q\phi_a)) \psipii =-p_\vel' c \psimii  \quad (b)\hfill
\cr
\hfill (\eng+(Mc^2-q\phi_a)) \psimi =p_\vel' c \psipi  \quad (c), \quad
(\eng+(Mc^2-q\phi_a)) \psimii =-p_\vel' c \psipii  \quad (d) \hfill (\ref{eq-flineq})
}$$
corresponding to (\ref{eq-matrx1}), for the wave functions.
Taking the imaginary of equations (a) and (b) first,   multiplying the resulting equation with equation (c)  and the second  with (d) respectively on opposite sides, 
substituting with  
$\psi_{\m }^*(1) \psimi
=C_{\m 1}^2$, 
$\psi_{\p }^* (1)\psi_{\p }(1)=C_{\p 1}^2$, 
$\psi_{\m }^*(2) \psi_{\m}(2)=C_{\m 2}^2$ and 
$\psi_{\p }^*(2) \psi_{\p }(2)=C_{\p 2}^2$, 
we get

\noindent
$
%$\displaylines{\refstepcounter{equation} \label{eq-D8b}\hfill
%\begin{array}{c}
(\eng -(Mc^2+q \phi_a)) C_{\p 1}^2 =  (\eng +Mc^2-q \phi_a)C_{\m 1}^2, \  
(\eng -(Mc^2+q \phi_a)) C_{\p 2}^2 
=  (\eng +Mc^2-q \phi_a)C_{\m 2}^2 
%\end{array}\hfill %\cr\hfill(\ref{eq-D8b})}$
$
\noindent
These have two independent solutions, and in mathematical terms  two of the four wave functions  can thus be arbitrarily 
chosen.
In view of the IED particle model by which each particle has internal, wave  processes consisting of two components  in the one-dimensional box, it is natural here that we choose the values for $C_{\p 1}$ and $C_{\p 2}$ symmetrically, in the sense also $C_{\p 2}= -C_{\p 1}$, with these in the two equations above           
                 %(\ref{eq-D8b}) 
the values for $C_{\m 1}$ and $C_{\m 2}$ then follow to be uniquely given as 
$$\displaylines{
\hfill
\begin{array}{c}
  {C_{\p 1} \atop C_{\p 2}} 
=\pm \lf(\sqrt{\frac{\eng +(Mc^2-q \phi_a)
             }{\eng -(Mc^2+q \phi_a)}} \rt)^{1/2} C,
\quad
 { C_{\m 1} \atop C_{\m 2}}
=\mp \lf(\sqrt{\frac{\eng -(Mc^2+q \phi_a)
                     }{\eng +(Mc^2-q \phi_a)}} \rt)^{1/2} C
\end{array}\hfill
}$$
where $
|C_{\p 1} C_{\m 1}|=  C^2, 
|C_{\p 2} C_{\m 2}|=  C^2 $.
With the above in the trial functions, 
we get the complete solution for Dirac equation
$$\displaylines{
\hfill {\pmb{\psi}}
=\left(
\begin{array}{c}
\psipi  \cr
\psipii   \cr
\psimi \cr 
 \psimii
\end{array}
\right)
=\left(
\begin{array}{c}
\lf(\sqrt{        \frac{\eng +(Mc^2-q\phi_a)
                          }{\eng -(Mc^2+q\phi_a)}
                          } \rt)^{1/2} C e^{i(k_d z-\w t)}  
\cr
-\lf(\sqrt{        \frac{\eng +(Mc^2-q\phi_a)}{\eng -(Mc^2+q\phi_a)}
} \rt)^{1/2} C e^{i(k_d z+\w t)}    
\cr
-\lf(\sqrt{         \frac{\eng -(Mc^2+q\phi_a)
                         }{\eng +(Mc^2-q\phi_a)}
         } \rt)^{1/2} C e^{i(k_d z+\w t)}  
\cr 
\lf(\sqrt{\frac{\eng -(Mc^2+q\phi_a)}{\eng +(Mc^2-q\phi_a)}} \rt)^{1/2} C e^{i(k_d z-\w t)} 
\end{array}
\right).
\refstepcounter{equation} \label{eq-Dx1}
\hfill               
%\cr \hfill  
 (\ref{eq-Dx1})
}
$$
Agreeing with the wave functions directly based on IED particle model, 
$\psipi,\psimi$ are two opposite travelling component waves of particle 1, and $\psipii,\psimii$ of particle 2; in the meantime, the spin-up component waves of particles 1 and 2, $\psipi$ and $\psipii$ travel in opposite directions and similarly the spin-down component waves.

\end{appendix}

The author would like to thank scientist P.-I. Johansson for his  moral and funding support of the research, the Committee of the 5th International Symposium on Quantum Theory and Symmetries (QTS 5) for a grant covering the conference fee,  and the Swedish Research Council for a travel grant enabling the author to present this work at the QTS 5, Valladolid  and  the Swedish Institute of Space Physics for administrating the travel grant. The author would like to thank Professor H.-D. Doebner, Professor J. Goldin and
Professor V. Dobrev    for  valuable reading of this and the related papers, for their valuable discussion and suggestions for a more informative and adaptable   introduction to the particle model, and thank several distinguished Professors in Sweden for valuable reading of this and the related papers. The author would also like to acknowledge the interesting discussion of Professors J. Patera, M. Berry,   J. Gazeau, D. Schuch, L. Boyle, R. Picken, A. Bohm, M. Olmo 
and others at the QTS 5.


\begin{thebibliography}{10}
\bibitem{Propagate} Mechanical wave can not be propagated  in empty space. Here the "$E_{\empty *}$ propagated in empty space" should be understood as the field $E$ propagated in the dielectric vacuum but after compensated  for the damping in amplitude by the dielectric medium.
\end{thebibliography}
\end{document}